\documentclass[twocolumn,superscriptaddress,aps]{revtex4-2}
\usepackage{amsmath,amssymb,bm,graphicx,color,gensymb,bbold,appendix,hyperref}
\usepackage[usestackEOL]{stackengine}
\begin{document}
\title{Constraining the Parameter Space of a Quantum Spin Liquid Candidate in Applied Field with Iterative Optimization}

\author{William Steinhardt}
\affiliation{Department of Physics, Duke University, Durham, North Carolina, 27008, USA}
\author{Zhenzhong Shi}
\affiliation{Department of Physics, Duke University, Durham, North Carolina, 27008, USA}
\author{Anjana Samarakoon}
\affiliation{Shull-Wollan Center, Oak Ridge National Laboratory,Oak Ridge, Tennessee 37831, USA 27008, USA}
\affiliation{Neutron Scattering Division, Oak Ridge National Laboratory, Oak Ridge, Tennessee 37831, USA}
\author{Sachith Dissanayake}
\affiliation{Department of Physics, Duke University, Durham, North Carolina, 27008, USA}
\affiliation{Neutron Scattering Division, Oak Ridge National Laboratory, Oak Ridge, Tennessee 37831, USA}
\author{David Graf}
\affiliation{National High Magnetic Field Laboratory and Department of Physics, Florida State University, Tallahassee, Florida 32310, USA}
\author{Yaohua Liu}
\affiliation{Neutron Scattering Division, Oak Ridge National Laboratory, Oak Ridge, Tennessee 37831, USA}
\author{Wei Zhu}
\affiliation{Theoretical Division, T-4 \& CNLS, Los Alamos National Laboratory, Los Alamos, New Mexico 87545, USA}
\affiliation{Westlake Institute of Advanced Study, Hangzhou 310024, China}
\author{Casey Marjerrison}
\affiliation{Department of Physics, Duke University, Durham, North Carolina, 27008, USA}
\author{Cristian D. Batista}
\affiliation{Department of Physics and Astronomy, The University of Tennessee,
Knoxville, TN 37996, USA}
\affiliation{Neutron Scattering Division and Shull-Wollan Center, Oak Ridge National Laboratory, Oak Ridge, TN 37831, USA}
\author{Sara Haravifard}
\affiliation{Department of Physics, Duke University, Durham, North Carolina, 27008, USA}
\affiliation{Department of Mechanical Engineering and Materials Science, Duke University, Durham, North Carolina 27708, USA}
\date{\today}

\begin{abstract}
The quantum spin liquid (QSL) state is an exotic state of matter featuring a high degree of entanglement and lack of long-range magnetic order in the zero-temperature limit. The triangular antiferromagnet YbMgGaO$_4$ is a candidate QSL host, and precise determination of the Hamiltonian parameters is critical to understanding the nature of the possible ground states.  However, the presence of chemical disorder has made directly measuring these parameters challenging.  Here we report neutron scattering and magnetic susceptibility measurements covering a broad range of applied magnetic field at low temperature.  Our data shows a field-induced crossover in YbMgGaO$_4$, which we reproduce with complementary classical Monte Carlo and Density Matrix Renormalization Group simulations. Neutron scattering data above and below the crossover reveal a shift in scattering intensity from $M$ to $K$ points and, collectively, our measurements provide essential characteristics of the phase crossover that we employ to strictly constrain proposed magnetic Hamiltonian parameters despite the chemical disorder. Constrained exchange parameters further suggest the material's proximity to the QSL state in the clean limit.  More broadly, our approach demonstrates a means of pursuing QSL candidates where Hamiltonian parameters might otherwise be obscured by disorder.
\end{abstract}
\maketitle
\section{Introduction}
The search for an experimental realization of the quantum spin liquid (QSL) state has been central to condensed matter physics for decades\cite{savary2016quantum,wen2019experimental}.  Anderson first proposed triangular antiferromagnets to exhibit the QSL state via the resonating-valence-bond (RVB)\cite{anderson1973resonating}, and the hunt for promising triangular lattice candidates systems is still ongoing \cite{zhong2019strong,bordelon2019field,liu2018rare,ashtar2019reznal,sarkar2019quantum,PhysRevB.95.060402,PhysRevMaterials.2.044403,iizuka2020single,xing2019synthesis}, though QSLs have been proposed for a variety of other models and systems, such as the kagome (corner-sharing triangular), honeycomb, and pyrochlores\cite{savary2016quantum,wen2019experimental,gingras2014quantum}.  Indeed, recent years have offered an explosion in possible theoretical models for QSLs, and a wide range of new candidate material hosts, but few new experimental methods to detect them\cite{knolle2019field}.  With the lack of a ``smoking gun,'' a variety of complimentary experimental methods, such as magnetic susceptibility and neutron scattering, are needed to establish the possible presence of the QSL state.  Equally important, numerical techniques such as DMRG are required to validate interpretation of the data as evidence of a QSL. Thus the parameters describing the magnetic Hamiltonian must be determined before the question of whether or not a material hosts the QSL ground state can be answered. For very recent examples, consider work on the exactly-solvable Kitaev system $\alpha$-RuCl$_3$\cite{suzuki2020quantifying}, or on the hyperkagome PbCuTe$_2$O$_6$\cite{chillal2020evidence}, or in the study of the triangular antiferromagnet YbMgGaO$_4$\cite{zhang2018hierarchy}, the focus of this work.  

Initial interest in YbMgGaO$_4$ (see Supplementary Figure 1 for crystal structure) stemmed from the absence of long range order at low temperatures\cite{li2015gapless}, in addition to the identification of strong spin-orbit coupling and an odd number of electrons per unit cell, as such features may contribute to exotic ground states\cite{savary2016quantum}.  Consequently, a multitude of theoretical\cite{zhu2017disorder,liu2016semiclassical,kimchi2018valence,wu2019randomness, zhu2018topography,parker2018finite,li2018effect,luo2018spinon,iaconis2018spin,li2017spinon,maksimov2019anisotropic,lima2019spatial, li2017detecting,gong2017global,luo2017ground,li2020reinvestigation,wu2020exact} and experimental\cite{ma2018spin,li2019rearrangement,toth2017strong,li2015rare,xu2016absence,li2017nearest,li2017crystalline,paddison2017continuous,shen2018fractionalized, li2016muon, zhang2018hierarchy,bachus2020field,ding2020persistent} studies sought to probe the low temperature physics and phase diagram, and elucidate the nature of the spin liquid-like phenomena.  A number of related systems have garnered significant attention\cite{cevallos2018structural,xu2019structure,shen2019intertwined, liu2018rare,ashtar2019reznal}. 

Efforts to understand the impact of chemical disorder due to site-mixing between non-magnetic Mg$^{2+}$ and Ga$^{3+}$ have enriched the discussions\cite{zhu2017disorder,kimchi2018valence,wu2019randomness,li2017nearest,li2017crystalline,li2015gapless,xu2016absence}. Inelastic neutron scattering measurements suggested that crystalline electric field (CEF) levels are broadened by a distribution of ytterbium-oxygen bond distances due to the Mg$^{2+}$/Ga$^{3+}$ site mixing, leading to a distribution of effective spin-half $g$-factors and consequently broadened low-energy magnetic excitations in the polarized state\cite{li2017crystalline}. Theoretical work suggests that, barring disorder in the charge environment from Mg$^{2+}$/Ga$^{3+}$ site mixing, YbMgGaO$_4$ should have a collinear/stripe ground state\cite{zhu2017disorder,parker2018finite}, while other calculations have suggested that the system could exhibit either a striped or 120$\degree$ ordered state\cite{gong2017global}.  While ac susceptibility measurements down to 60 mK imply spin freezing\cite{ma2018spin}, inelastic neutron scattering suggests roughly 16$\%$ of spins frozen\cite{paddison2017continuous}, and dc susceptibility measurements provide evidence of dynamic spins down to 40 mK with only approximately 8$\%$ frozen\cite{li2019rearrangement}.  The low fraction of frozen spins measured in Ref. \citenum{paddison2017continuous} has previously been attributed to a high degree of frustration\cite{ma2018spin}.  Heat capacity measurements\cite{li2015gapless} and $\mu$SR studies\cite{li2016muon} at 60 mK similarly oppose a spin freezing scenario.  Very recently, $\mu$SR studies were carried out down to temperatures as low as 22 mK without observing ordered or disordered static magnetism \cite{ding2020persistent}, further opposing  interpretations of the ground state as a spin glass.

\begin{figure*}
\includegraphics[width=\linewidth]{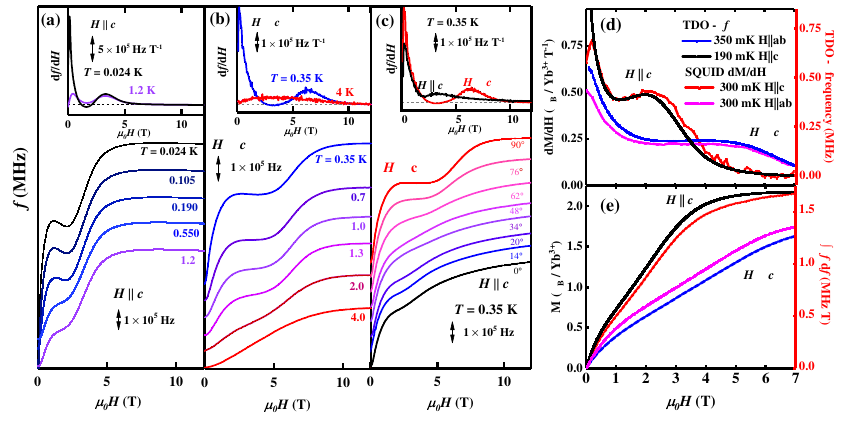}
\vskip -0.2cm
\caption{Tunnel diode oscillation (TDO) frequency and cantilever torque response versus applied field. (a) and (b) TDO ($\Delta f/f \propto \Delta \textbf{M}/ \Delta \textbf{H}$) shows an anomaly for $\textbf{H}\parallel c$ and  $\textbf{H}\perp\textbf{c}$ respectively which weakens as temperature increases (curves offset for clarity in (a) (b)). (c) Anomaly’s response to applied field shows anisotropy.  (e) $d\textbf{M}/d\textbf{H}$ measured with SQUID corroborates TDO measurement. (f) integrating TDO $\Delta$frequency from 0 to the approximate saturation is further corroborated by dc susceptibility measurements (See Materials and Methods in Supplementary Materials. } 
\label{TDO_SQUID}
\end{figure*}

Precisely measured exchange parameters are needed to evaluate the possibility of a QSL state.  Conventionally, in systems with complex Hamiltonians, exchange parameters are extracted from fits of the single-magnon dispersion obtained from inelastic neutron scattering.  Quenched disorder presents a significant obstacle to this endeavor, introducing broadening observed in the dispersion in YbMgGaO$_4$\cite{paddison2017continuous}.  Though a timed-resolved THz spectroscopy study significantly refined constraints\cite{zhang2018hierarchy}, difficulties are compounded by the distribution of effective $g$-factors\cite{li2018effect}. Furthermore, the bond-dependent $J_{z\pm}$ exchange parameter, important in the ground states at fields below saturation, is absent from the analytical expression of the dispersion in the polarized state obtained from the linear spin wave approximation\cite{li2018effect}.  In order to properly evaluate proposed Hamiltonian parameters, systematic measurements at low and intermediate field are required.  More generally, magnetic field is an effective and extensively used tuning parameter in the study of QSL candidates\cite{ma2018recent}, such as to suppress a competing ordered state\cite{kasahara2018majorana}, or to suppress spin fluctuations and measure spin waves in high field\cite{paddison2017continuous}, and a number of recent triangular antiferromagnetic materials show interesting phase transitions or features in applied field\cite{xing2019field,xing2019synthesis}. A detailed study of YbMgGaO$_4$’s field dependence, including at low and intermediate fields, is a natural next step to understand the ground states of the system, and to provide further data for comparison to proposed models.

To that end, this work presents new data revealing the field evolution of YbMgGaO$_4$, including a magnetic phase crossover, and by establishing relevant criteria which we employ to systematically constrain the possible Hamiltonian parameters.  We conducted measurements of high-quality, single crystalline YbMgGaO$_4$ using the ultra-sensitive tunnel diode oscillator technique \cite{shi2019emergent} (TDO), magnetization via SQUID, cantilever torque magnetometry, and diffuse neutron scattering.  As a function of temperature and applied field, varying sample orientations using TDO, SQUID, and cantilever torque magnetometry measurements are used to show anisotropy of the bulk response, while diffuse neutron scattering provides the finite $\textbf{Q}$ dependence of spin correlations.  All provide clear evidence of a field-induced phase crossover, which must be reproduced by any set of Hamiltonian parameters describing the magnetic ground states and phase diagram of YbMgGaO$_4$. We compare our experimental data to complementary Monte Carlo (MC) and Density Matrix Renormalization Group (DMRG) calculations, where the results demonstrate that despite the diverse range of exchange parameters proposed for this system \cite{li2018effect,paddison2017continuous,li2015rare,zhang2018hierarchy,toth2017strong}, only strictly constrained values can reproduce the field-induced phase crossover observed. We further employ a methodical and systematic approach to reduce the possible volume in the 7-dimensional parameter space.  These results further elucidate the effects of disordered exchange interactions and $g$-factors and suggest a magnetic field-induced phase transition in the disorder-free limit.  Furthermore, the strictly constrained parameter set suggests that YbMgGaO$_4$’s ground state is proximal to the QSL state in the disorder-free limit predicted by \cite{zhu2018topography}, as will be discussed below. 

\section{Methods}
\subsection{Sample Synthesis and Characterization}
Powder of YbMgGaO$_4$ (structure shown in Supplementary Figure 1) was produced by finely grinding mixed powders of Yb$_2$O$_3$, MgO, and Ga$_2$O$_3$, and reacting at 1150$\degree$ C in a box furnace.  The product was ground again to a fine powder, compressed hydrostatically into a rod, and then sintered at 1500$\degree$ C in a vertical Bridgman furnace.  Finally, large single crystals of YbMgGaO$_4$ were grown using the optical floating zone method (Supplementary Figure 2).  A typical crystal was grown in an O$_2$ atmosphere at 1 MPa, with an initial growth speed of 20 mm/hr, and upon stabilization of the liquid zone, ~ 4 mm/hr until finished.  

We confirmed the powder and crystal phase at each step of the synthesis using ground powders and the Panalytical X’Pert PRO MRD HR XRD System (using Cu K-$\alpha$ 1.5418 nm X-rays) and Rietfeld refinement via FULLPROF\cite{rodriguez1993recent}.  Single-crystal quality and alignment was determined using Laue X-ray diffraction (Supplementary Figure 3).  Preliminary susceptibility and measurements were carried out on a powder sample using a Quantum Design MPMS XL7 SQUID magnetometer down to 1.8 K, and confirmed previous measurements of the Curie-Weiss temperature $\Theta \approx -4$ K.

\subsection{High Sensitivity Magnetization Measurements}
High-sensitivity measurements of magnetization were achieved with the complimentary tunnel diode oscillator (TDO) technique (Fig. 1) and torque magnetometry. In a TDO measurement, a tunnel diode is biased to operate in the “negative resistance” region of the IV-curve. This provides power that maintains the resonance of a LC-circuit at a frequency range between 10 and 50 MHz. An approximately cylindrical single-crystal sample with dimensions of ~2 mm in length and ~1 mm in diameter was placed inside a detection coil, with the $\textbf{c}$ axis of the sample aligned with the coil axis (Supplementary Figure 5). Together, they form the inductive component of the LC circuit. Changes in sample magnetization induce a change in the inductance, hence a shift in the resonance frequency. Highly sensitive measurements in changes of magnetic moments ~ $10^{-12}$ e.m.u., therefore, are enabled by the ability of measuring the resonance frequency to a high precision\cite{van1975tunnel}.  The magnetization and susceptibility results collected at temperatures down to $T =$ 1.8 K (Supplementary Figure 4) were consistent with earlier reported results \cite{paddison2017continuous,han2012fractionalized,li2015rare}.  

Magnetization was also directly measured via an in-house Cryogenic S700X SQUID magnetometer in temperatures down to 280 to 300 mK using a Helium 3 probe.  A 1.3 mg sample was mounted on a silver straw with vacuum grease in $\textbf{H} \parallel \textbf{c}$ and $\textbf{H}\perp\textbf{c}$ orientations.  See Fig. 1d and 1e.   

\subsection{Neutron Scattering}
Neutron scattering data was collected at the CORELLI spectrometer at Spallation Neutron Source, Oak Ridge National Laboratory\cite{rosenkranz2008corelli}. CORELLI is a quasi-Laue TOF instrument equipped with a large 2D detector, with a -20$\degree$ to +150$\degree$ in-plane coverage and $\pm$28.5$\degree$ out-of-plane coverage. The incident neutron energy was between 10 meV and 200 meV. A superconducting magnet was used to provide a vertical magnetic field up to 5 T, which reduced the effective out-of-plane coverage to $\pm$ 8$\degree$. An 896.10 mg single crystal was mounted on a Cu plate in a dilution refrigerator. The sample was aligned with the $(\textbf{h}, \textbf{k}, 0)$ plane horizontal and the magnetic field along the [0,0,$\textbf{l}$] direction. Neutron-absorbing Cd was used to shield the sample holder to reduce the background scattering. Experiments were conducted with applied fields at the base temperature 130 mK by rotating the crystal through 180$\degree$ in 3$\degree$ steps, and then at 20 K in the same fields for background subtraction. The data were reduced using Mantid for the Lorentz and spectrum corrections\cite{michels2016expanding}.

To ensure our observations utilizing total scattering mode were consistent with measurement at the best possible energy resolution of the instrument, we also measured 130 mK and 20 K at 0, 3, and 5 T with the correlation chopper on (see supplementary figure 8).  We further filtered the TOF data to provide an estimated energy resolution of ~0.2 meV and identified features completely consistent with our measurement in total scattering mode.  

To account for the temperature factor in our background subtraction in total-scattering mode, we compared the ratio of integrated intensities of a small region in reciprocal space at 5 T for both temperatures.  The region was bounded by $-0.9 < \textbf{h} < -0.75$, $0.75 < \textbf{k} < 1.1$, and $-0.5 < \textbf{l} < 0.5$ (the $ \textbf{l}$ range is consistent with our BZ edge and planar integrations).  We used the second BZ and 5 T data to avoid skewing the integration at low temperature due to diffuse magnetic scattering.  This integration approximates the ratio of Bose population factors - our scaling factor was $1.133\pm0.005$. To improve statistics, we used the 6-fold symmetry (in accordance to the symmetry of the $\textbf{l}$ = 0 plane) to add our data 6 times at subsequent 60$\degree$ angles and average.  All analysis and visualization were performed using Mantid and Python.

\section{Results}
We performed ultra-highly sensitive magnetic susceptibility measurement on YbMgGaO4 using TDO technique (Supplementary Figure 5) up to fields exceeding saturation and at temperatures down to 24 mK (Fig. 1).  Comparison to magnetization measured at 300 mK in SQUID (Fig. 1e, f) and cantilever torque magnetometry (Supplementary Figures 5 and 6) results confirms the appearance of an anomaly near 2 T. The anomaly weakens with increasing temperature, beginning to soften after 190 mK and vanishing near 4 K (Fig. 1b,c).  The greater prominence observed in both techniques with applied field perpendicular (versus parallel) to the $\textbf{c}$-axis (Fig. 1b) is consistent with easy-plane anisotropy in agreement with earlier experiments\cite{li2015rare}.  The anomaly is likely a remnant of the quantum mechanical 1/3 magnetization plateau\cite{chubukov1991quantum}, as will be discussed below.
\begin{figure}[t]
\includegraphics[width=\linewidth]{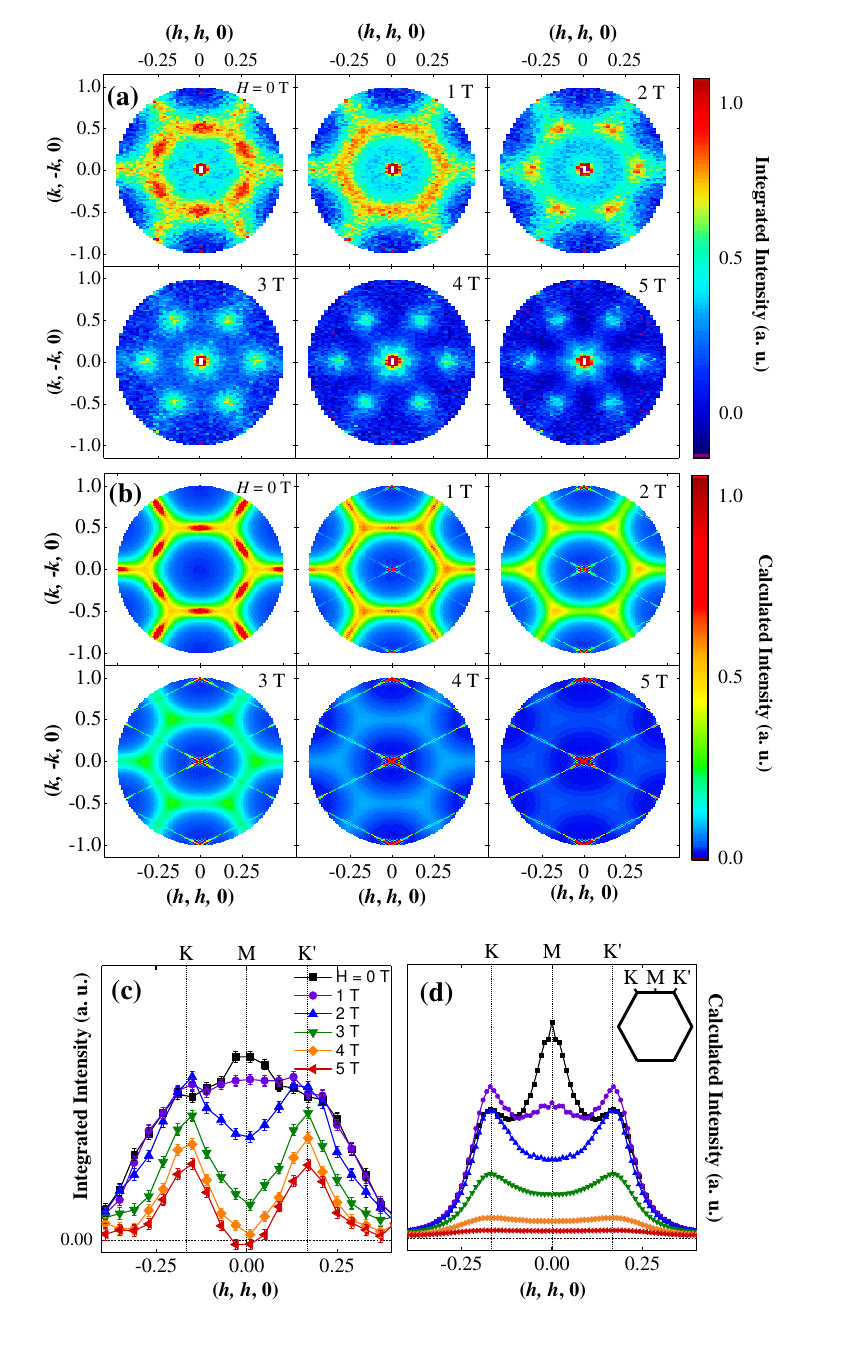}
\vskip -0.5cm
\caption{Diffuse neutron scattering from magnetic structure for a series of fields with $\textbf{H} \parallel \textbf{c}$. (a) Color maps of Brillouin zone integrated $(-0.5 < \textbf{l} < 0.5)$.  As field increases, spectral weight shifts from $M$ points to $K$ points, then $\Gamma$ points, in agreement with (b) calculated $S(\textbf{Q})$ with $\sigma_{g_\perp} =0.3$, $\sigma_{g_{\parallel}} =1.2$, $\sigma_J=0.5$  at 130 mK.  (c) Integrated neutron scattering intensity $(-0.5 < \textbf{l} < 0.5)$ along $(\textbf{h},\textbf{h},0)$ direction. (d) Calculated $S(\textbf{Q})$ along $(\textbf{h},\textbf{h},0)$.}
\label{neutron}
\vskip -0.4cm
\end{figure}

To gain insight into the changing magnetic structure corresponding to the anomaly detected in the  magnetometry measurements, and to exploit the advantages offered by neutron scattering in revealing spin correlations at finite $\textbf{Q}$, we employed diffuse neutron scattering in total scattering mode at CORELLI, an instrument optimized for measuring short-range correlations.  Data was collected with the sample $\textbf{c}$ axis parallel to the applied field direction (Fig. 2a).  20 K background signal was subtracted from 130 mK data for integer fields from $0$ to $5$ T, isolating the magnetic contribution. To improve statistics, we averaged intensity about the $\textbf{l}$-axis by applying allowed symmetry operations (see Supplementary Materials for details). Integration over $-0.5 < \textbf{l} < 0.5$ in reciprocal space for zero field shows diffuse magnetic scattering centered at high-symmetry $M$ points, indicative of short-range correlations and consistent with previous neutron measurements\cite{paddison2017continuous,han2012fractionalized,toth2017strong}.  However, we discovered that with increasing applied field the spectral weight at the $M$ points subsides.  At $\mu_0\textbf{H} = 1$ T the spectral weight is flat along the zone edge, while at $\mu_0\textbf{H} = 2$ T most of the diffuse spectral weight is distributed at the $K$ points, coinciding with the feature in our magnetic susceptibility data (Fig. 1).  Importantly, we note that the intensity at the $K$ point below 2 T is mostly independent of the applied field.  When field is further increased above 2 T, the scattering intensity at $K$ points diminishes as the system approaches the polarized state. The field dependence for $2$ to 5 T is consistent with the gradual saturation of the curve observed in our magnetic susceptibility data. Integration of the spectral weight in a narrow rectangular volume along the first Brillouin Zone (BZ) edge shows the shift in spectral weight from the $M$ to $K$ points (Fig. 2c).  This shift in intensity indicates a change in short-range correlations and underlying spin structure, and likely relates to a crossover from a stripe-type state to 120$\degree$-type state in the clean limit.  As field is further increased, scattering intensity along the zone edge is reduced as seen in Fig. 2a. As diffuse scattering measurements have a finite energy resolution about the elastic line, we corroborated our results measured in total scattering mode using a correlation chopper and time-of-flight (TOF) filtering to achieve an energy resolution of $E \approx 0.2$ meV (see Supplementary Figure 8).

\begin{figure}[t]
\includegraphics[width=\linewidth]{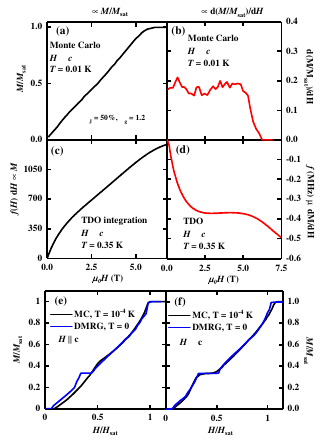}
\caption{Calculated magnetization curves and background-subtracted integrated TDO measurement for $\textbf{H}\perp \textbf{c}$ and magnetization curves obtained with Monte Carlo sampling and DMRG . (a) Classical Monte Carlo simulation of magnetization (featuring bi-quadratic terms and disorder) at $T = 0.01$ K.  (b) Derivative with respect to applied field of classical Monte Carlo simulation in (a). (c) Integration the change in TDO frequency at $T = 0.35$ K. (d) Change in TDO frequency at $T=0.35$ K. (e) Comparison of magnetization from DMRG calculations (red) versus classical simulations with bi-quadratic interaction ($\lambda_1=0.3$, $\lambda_2=0.03$) when $\textbf{H} \parallel \textbf{c}$ and (f) $\textbf{H}\perp\textbf{c}$.  Addition of bi-quadratic terms mimics quantum mechanical plateau for applied field perpendicular to sample $c$ axis.}
\label{magnetization}
\vskip -0.4cm
\end{figure}

To shed light on the nature of the observed phenomena, we start by considering a classical limit of the spin Hamiltonian proposed in previous works \cite{li2016anisotropic}:
\begin{multline}
{\cal H} = \sum_{\langle ij \rangle} \Big[J_{zz}^{(1)} S_i^z S_j^z + J_{\pm}^{(1)} (S_i^+ S_j^- + S_i^- S_j^+) \\ + J_{\pm\pm}^{(1)}(\gamma_{ij}S_i^+S_j^+ + \gamma_{ij}^* S_i^-S_j^-)\\ - \frac{iJ_{z\pm}^{(1)}}{2}(\gamma_{ij}^*S_i^+S_j^z - \gamma_{ij} S_i^-S_j^z + \langle i \leftrightarrow j \rangle)\Big] \\ + \sum_{\langle\langle ij \rangle\rangle} \Big[J_{zz}^{(2)} S_i^z S_j^z + J_{\pm}^{(2)} (S_i^+ S_j^- + S_i^- S_j^+)\Big]
\end{multline}  
with  phase factors $\gamma_{ij}=1$, $e^{i2\pi/3}$, $e^{-i2\pi/3}$ for each of the three principal directions of the triangular lattice. The bond notation $\langle ij \rangle$ ($\langle\langle ij \rangle\rangle$) indicates that the corresponding sum runs over pairs of nearest (second-nearest) neighbors. The corresponding exchange nearest-neighbor and second nearest-neighbor exchange tensors are,
\begin{multline}
J_{ij}^{(1)} =
\begin{bmatrix}
2J_{\pm}^{(1)}+2J_{\pm\pm}^{(1)}\tilde{c}_\gamma & -2J_{\pm\pm}^{(1)} \tilde{c}_\gamma & -J_{z\pm}^{(1)} \tilde{s}_\gamma \\
-2J_{\pm\pm}^{(1)} \tilde{s}_\gamma & 2J_{\pm}^{(1)}-2J_{\pm\pm}^{(1)} \tilde{c}_\gamma & J_{z\pm}^{(1)} \tilde{c}_\gamma \\
-J_{z\pm}^{(1)} \tilde{s}_\gamma & J_{z\pm}^{(1)}\tilde{c}_\gamma & J_{zz}^{(1)}
\end{bmatrix},
\nonumber
\end{multline}
and
\begin{equation}
J_{ij}^{(2)} =
\begin{bmatrix}
2J_{\pm}^{(2)} & 0 & 0 \\
0 & 2J_{\pm}^{(2)} & 0 \\
0 & 0 & J_{zz}^{(2)}
\end{bmatrix},
\nonumber 
\end{equation}
respectively, where $\tilde{c}(\tilde{s})_\gamma\!=\!\cos\tilde{\theta}_\gamma (\sin\tilde{\theta}_\gamma)$ and $\gamma$ indexes the three directions.

In a first attempt, we manually optimized the Hamiltonian parameters starting with a potential set given in the Ref.~\citenum{li2018effect} and  exploring the parameter space to capture details of the field dependent behavior and single magnon dispersion.  We considered the set of Hamiltonian parameters, $J_\pm^{(1)}=0.66 J_{zz}^1$, $J_{\pm\pm}^{(1)}=0$,  $J_{z\pm}^1=0.13 J_{zz}^{(1)}$, $J_{zz}^{(2)}=0.1 J_{zz}^{(1)}$, $J_{\pm}^{(2)}=0.066 J_{zz}^{(1)}$ and $J_{zz}^{(1)}=0.164$ meV, that reproduce the phenomenology better (featured as the magenta star in Figure 4). For the average $g$-factor, we used $g_{\parallel}=3.7227$\cite{toth2017strong}.  

To quantify the uncertainty of the proposed model Hamiltonian for a clean version of YbMgGaO4, we applied the iterative optimization procedure explained in Ref. \citenum{samarakoon2019machine}.  The experimental input includes the magnon dispersion at 7.8 T measured with Inelastic Neutron Scattering (INS)\cite{paddison2017continuous} and the field-induced crossover revealed by our magnetic diffuse scattering measurements. The challenge resides in the combination of a high-dimensional (d=7) Hamiltonian space ($\cal H$ includes seven independent parameters $J_\pm^{(1)}$, $J_{\pm\pm}^{(1)}$, $J_{z\pm}^{(1)}$, $J_{zz}^{(1)}$, $J_\pm^{(2)}$, $J_{zz}^{(2)}$ and $g_{\parallel}$),  and the significant amount of disorder present in YbMgGaO$_4$. Given that both sources of complexity are present in many other frustrated materials with large spin-orbit coupling, it is important to develop new protocols to extract models from data and simultaneously estimate their uncertainty.

We then implemented an optimization protocol in three steps. First, we
introduced the the cost-function:
\begin{equation}
\chi_{INS}^2 = \sum_{\textbf{Q}=path_1+path_2} \Big( I_{peak}(\textbf{Q})-I_{cal.}(\textbf{Q}) \Big)^2,    
\end{equation}
to fit the measured magnon dispersion at $\mu_0 H =7.8$ T with the analytical expression obtained from linear spin wave theory (see Ref. \cite{paddison2017continuous}) along two reciprocal space pathways (see Supplementary Figure 9). 

As explained in Ref. \citenum{samarakoon2019machine}, for each iteration we used random samples over the whole Hamiltonian space to build a low-cost estimator of $\chi_{INS}^2$. We then used $\hat{\chi}_{INS}^2$ to evaluate the next set of Hamiltonian parameters uniformly distributed over the Hamiltonian space and subjected to the constraint $\hat{\chi}_{INS}^2 < c$. The cutoff $c$ is lowered after each iteration. The last iteration is reached when $c$ reaches its final value, $c_{final}$, for which the calculated dispersion agrees with the INS data within the experimental uncertainty. We note that experimental uncertainty is higher than the instrument resolution because the chemical disorder broadens the single-magnon line-shape. The manifold of model Hamiltonians that fit the measured spin-wave dispersion is indicated by the blue contours in Fig.~\ref{magnetization}.  

Second, we performed classical Monte-Carlo simulations (see Methods: Simulation) on the disorder-free model within the aforementioned manifold. For a given Hamiltonian space sample, we computed the dynamical spin structure factor:
\begin{equation}
S(\textbf{Q}, \omega) = \sum_{\alpha,\beta} \frac{g_{\alpha}g_{\beta}}{4} \left(\delta_{\alpha\beta}-
\frac{q_{\alpha}q_{\beta}}{q^2}\right)
|F(\textbf{Q})|^2
\mathcal{S}^{\alpha\beta}\left( \textbf{Q}, \omega \right)
\label{eq:neutron_scattering}
\end{equation}
where \textbf{Q} is the wavevector in the scattering process, $\alpha,\beta=x,y,z$ are cartesian coordinates indicating initial and final spin polarization of the neutron, $F(\textbf{Q})$ is the magnetic form factor and $\mathcal{S}^{\alpha\beta}\left( \textbf{Q},\omega \right)$ is the Fourier transform of the two-point spin correlation function. Note that for a classical spin model, $S(\textbf{Q}, \omega=0) \simeq S(\textbf{Q}) \equiv \int_0^{\infty} S(\textbf{Q}, \omega) d \omega$ at low enough temperature.  The MC simulations are performed at $T = 130$ mK for three values of the magnetic field (0 T, 1 T and 2 T). We then applied the constraint that the M-peak must be dominant below $\mu_0 H=1$ T, while the $K$-peak should become dominant at $\mu_0 H = 2$ T ($\textbf{H} \parallel \textbf{c}$ axis). The resulting submanifold of possible model Hamiltonians is delimited by the green color contour in Fig.~\ref{magnetization}. Finally, to further reduce the model uncertainty, we introduced another cost-function,

$$ \chi_k^2 = \sum_{\textbf{H}=0, 1 T, 2 {T}} \Big( I_K^{exp}(\textbf{H}) - I_K^{sim}(\textbf{H}) \Big)^2 $$
where $I_K$ is the integrated intensity under the $K$-peak, and applied the same optimization scheme to the MC calculations. We chose the $K$-peak intensity because our MC simulations of the disordered model indicate that it is less-susceptible to chemical disorder in comparison to the $M$-peak intensity (see Supplementary Figure 14).  

The region constrained by the condition $\chi_{combined}^2=\chi_{INS}^2+\chi_k^2<c^k$ is shown by the magenta color contours in Fig.~\ref{magnetization}. Due to our limited knowledge of the chemical disorder in this material, it is not possible to set a definite value for $c_{final}^k$. Consequently, we are presenting three solid, dashed and dotted contour lines, corresponding to higher to lower values of $c_{final}^k$, that reveal the landscape of $\chi_{combined}^2$. We note that the set of parameters that we found by manually optimizing the Hamiltonian parameters, indicated with a star in Fig.~\ref{magnetization}, lies within the range of possible solutions.

As we discuss below, the disorder present in YbMgGaO$_4$ introduces a significant variation in the value of these Hamiltonian parameters. To demonstrate this statement, we introduce chemical disorder in the exchange interactions ($J$-disorder) and in the $g$-factors ($g$-disorder) and assume that both have a uniform distribution of predefined width around the aforementioned mean values, $g_{\parallel i}= g_{\parallel_{av}}+\Delta_i (\sigma_{g_\parallel})$, and $J_{ij} = J_{av} (1+\Delta_{ij} (\sigma_J ))$, where  $\sigma_{g_\parallel}=1.2$ and $\sigma_J=0.5$.  The static magnetic structure factor $S(\textbf{Q})$ is calculated by a standard Metropolis sampling algorithm while the dynamical structure factor $S(\textbf{Q},\omega)$ is computed by Landau-Lifshitz dynamics\cite{samarakoon2017comprehensive}. Both types of structure factors shown throughout this paper are calculated by averaging over 60 independent sets of 2000 configuration samples from a standard Metropolis sampling algorithm on a 48 $\times$ 48 supercell (2304 spins), followed by a slow annealing process starting from a random disordered spin configuration. 

In Fig. 2 we show the comparison between experimental data and the theoretical simulations throughout the field-induced phase crossover, where main aspects of the experimentally observed phenomenology can be reproduced. This agreement includes the evolution of the diffuse scattering and the non-linearity observed in the magnetization curve at finite fields. Interestingly, only a strictly selective set of parameters can reproduce the experimental results for all fields and capture the crossover (see Fig. 5 and discussion below).  This is so despite the broadening of the observed magnon dispersion in the polarized state that is induced by the chemical disorder  (see Supplementary Figure 17).   
\begin{figure}[t]
\includegraphics[width=\linewidth]{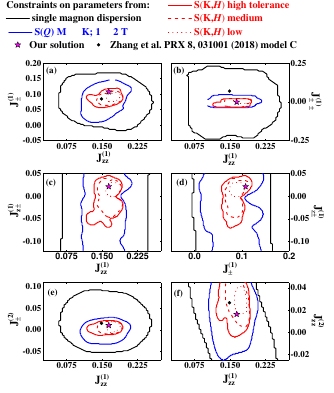}
\vskip -0.4cm
\caption{Contour plots of projected hyper volumes that result from different sets of experimental observations. Each sub panel is a 2D projection of 7D hyper-volumes. The uncertainties based on spin-wave dispersion (blue solid line), field induce transition (green solid line) and field evolution of integrated intensity of the $K$-peak $I_k (\textbf{Q})$(red solid line), are shown by multi-color contours. The contour red lines reveal the shape of the cost function $\chi_{combined}^2$ because they correspond to different values of the cutoff  $c^k$. Solid, dashed and dotted red lines correspond to the largest, intermediate and smallest value of $c^k$, respectively.}
\label{parameters}
\vskip -0.4cm
\end{figure}

To account for the effects of quantum fluctuations, we have computed the $\textbf{M}(\textbf{H})$ curve with the DMRG method on the $S=1/2$ version of $\cal H$ (without disorder) for field directions parallel to the $\textbf{c}$-axis and to the ab-plane. The resulting $\textbf{M}(\textbf{H})$  curve exhibits a characteristic plateau at 1/3 of the saturation value\cite{chubukov1991quantum,kamiya2018nature}, which is bigger for $\textbf{H}\perp\textbf{c}$ because of the easy-plane anisotropy (Fig. 3). This plateau phase is a true quantum mechanical signature where quantum fluctuations favor collinear configurations. The effect of quantum fluctuations can then be reproduced by adding an effective bi-quadratic interaction to the classical Hamiltonian\cite{kamiya2018nature,nikuni1998quantum}
\begin{equation}
 {\cal H}_1 = {\cal H} - \lambda_1 \sum_{\langle ij \rangle} \Big( S_i\cdot S_j \Big)^2 - \lambda_2 \sum_{\langle \langle ij \rangle \rangle}\Big( S_i\cdot S_j \Big)^2, 
 \end{equation}
where ${\cal H}$ is the classical spin Hamiltonian given in Eq. (1) and $\lambda_1=0.3J^{(1)}_\pm$ and $\lambda_2= 0.03 J^{(1)}_\pm$ are effective bi-quadratic couplings for next and second nearest-neighbor spins, respectively. The values of  $\lambda_1$ and $\lambda_2$ have been obtained by comparing the $M(H)$  curve obtained with DMRG for the $S=1/2$ version of ${\cal H}$ and the ${M}(H)$ curve obtained from a classical MC simulation of ${\cal H}_1$. 
To compare the calculations and TDO magnetic susceptibility results, we compare the change in resonant frequency (directly proportional to $dM/dH$ measured with SQUID – see Fig. 1d,e) then integrate with respect to applied field to obtain a curve proportional to magnetization (Fig. 3c), revealing a distinct non-linearity. This non-linearity can also be seen in the low-temperature calculation result (Fig. 3a), where disorder has been included. These results suggest that the suppression of the feature in $M(H)$  curves reported in earlier work \cite{paddison2017continuous,han2012fractionalized,li2015rare} could be due to thermal fluctuations (since these were measured from 1.7 K to 2 K). Furthermore, we note that the lower 0.5 K curve measured from a powder sample as reported in Ref. \citenum{li2015gapless} captures the field induced anomaly and agrees with our results at comparable temperatures.  Lower temperature and more sensitive techniques are required to detect the anomaly with better resolution. 

A comparison of the calculated $S(\textbf{Q})$  in the absence of disorder at $T = 0$ (Supplementary Figure 18) to $T=130$ mK (Supplementary Figures 19) shows that thermal fluctuations are enough to disrupt long-range order, owing to the high degree of frustration.  Furthermore, the calculated $S(\textbf{Q})$ in the presence of disorder as $T$ approaches zero also shows disruption of long-range order (Supplementary Figure 21). These findings indicate that the lack of magnetic Bragg peaks and consequently the broad continuum of magnetic excitations can likely be attributed  to  thermal fluctuations or chemical disorder. In other words, even if it were possible to produce cleaner samples of YbMgGaO$_4$, our classical analysis indicates that lower temperatures would be necessary to measure the presence of long range order. While this observation does not preclude the possibility of a spin liquid state induced by quantum fluctuations, it indicates that the lack of Bragg peaks down to $T=130$ mK cannot be regarded as a strong indicator of spin liquid behavior.

Recent theoretical studies of the effect of disorder on the $J_1-J_2$ Heisenberg triangular antiferromagnets show that disorder may contribute more intensity along the BZ edge (in zero field)\cite{wu2019randomness}.  Likewise, as noted above, chemical disorder is essential to reproduce the broadening of the magnon peaks (Supplementary Figure 17) at $\mu_0 H = 7.8$ T as also observed in experiment\cite{paddison2017continuous}. We therefore suggest that both thermal fluctuations and chemical disorder  contribute to the observed lack of a long-range order in YbMgGaO$_4$. In particular, we note that as QSL candidates in general are highly frustrated, they may be similarly susceptible to the effects of thermal fluctuations (even at very low temperatures) and thus these effects should be taken into consideration in future studies of this compound and other highly frustrated  systems with relatively small Curie-Weiss temperatures.

The above observations leave us with the following question: how can we use the available experimental information to diagnose quantum spin liquid behavior in the clean limit? One possible approach is to extract a model Hamiltonian for the clean limit of YbMgGaO$_4$ and compare the result against the quantum phase diagram that is obtained from numerical techniques, such as DMRG~\cite{zhu2018topography}, that can account for non-perturbative effects of quantum fluctuations.
As we will briefly elaborate upon below, when we extend our work to optimize a case with isotropic $J_2$ interactions, we find parameters that lie in the spin liquid region indicated in Fig. 4 of Ref. \citenum{zhu2018topography}).

\begin{figure*}[t]
\includegraphics[width=\linewidth]{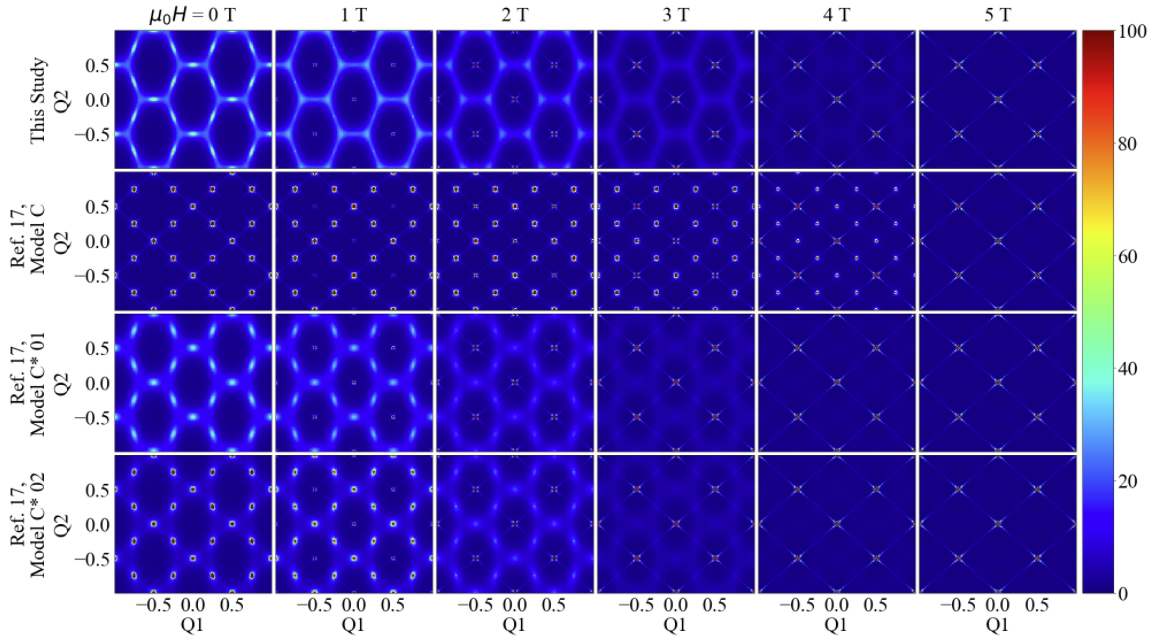}
\vskip -0.25cm
\caption{Comparison of field dependence of $S(\textbf{Q})$ calculated for parameters from four unique models, including those found in this study and as suggested by constraints determined in Ref. \citenum{zhang2018hierarchy}. $S(\textbf{Q})$ calculated for integer fields 0-5 T at 130 mK.  Note that the field-induced transition visible in the diffuse scattering data is absent for the parameters other than those used for this study, suggesting stringent constraints on the possible models for the system.}
\label{comparison}

\end{figure*}

\section{Discussion}
Definitively proving the existence of the quantum spin liquid ground states in candidate systems remains a critical experimental challenge\cite{wen2019experimental}.  The lack of long-range order in QSL states, coupled with the quenched disorder of many real experimental systems like YbMgGaO$_4$, suggests that future studies will benefit greatly from an improved understanding of the role disorder can play in QSL phenomena.  Indeed, disorder has been identified as a contributing factor for several proposed QSL candidates\cite{wen2017disordered,furukawa2015quantum}. However, it is important to recognize that a good characterization of the kind of disorder that is present in a specific compound  can be extremely challenging. If the minimal spin Hamiltonian of a given material has multiple parameters (seven for the case of YbMgGaO$_4$), extracting an accurate  model Hamiltonian is already very challenging in the clean limit. 

 An alternative approach is to study whether a system with quenched disorder has a QSL state in the \emph{ disorder-free limit} to determine if it is worth investigating  ways of reducing the level of chemical disorder in a specific material.
 The present work proceeds accordingly: the presence and unique characteristics of a phase crossover have been used to narrowly constrain the possible range of  magnetic Hamiltonian parameters, so that the Hamiltonian in the clean-limit can be established and the effects of disorder interpreted with that in mind.  We emphasize that despite the high quality of past studies, more information about the low and intermediate field behavior are critical to map out the low temperature magnetic phase diagram.  Our observation of the crossover in YbMgGaO$_4$ yields a lower-field reference point for comparison to measurements of the polarized state.  Furthermore, our results show a remarkably high sensitivity of the field-induced crossover to Hamiltonian parameters.  As an example, we consider recent time-domain THz spectroscopy (TDTS) measurements employed in conjunction with inelastic neutron scattering to measure the disorder-averaged exchange interactions in YbMgGaO$_4$\cite{zhang2018hierarchy}.  This work found a slightly larger $g$-factor $(g_{\parallel}=3.81)$ than previously reported.  We applied the exchange parameters and $g$-factor proposed in model C of Ref.~\citenum{zhang2018hierarchy} as well as a few variants of this model. Remarkably, we find that these parameters do not reproduce either the diffuse scattering, or the field-induced crossover we observe (Fig.~\ref{comparison}) – this is despite the close agreement between the $S(\textbf{Q}, \omega)$ spin waves at higher field calculated with both sets of parameters (Supplementary Figure 22).  The sensitivity of the phase crossover to the parameters is exemplified by Fig.~\ref{magnetization} b, where only a small discrepancy in the value of $J_{\pm\pm}$ contributes to the absence of the crossover in applied field (see Fig.~\ref{comparison}).  However, we note that optimizing the $g$-factor improves the quality of ours fits and yields a value of $g_{\parallel}=3.84$, in good agreement with Ref. \citenum{zhang2018hierarchy} (see supplementary figure 13).

\begin{table*}[]
\centering
\begin{tabular}{| c  c  c  c  c  c  c  c c |}
\hline
Parameter    & \Centerstack{Our\\Solution} & \Centerstack{Center of \\ Solution} & Uncertainty & \Centerstack{Ref. \citenum{zhang2018hierarchy} \\ Model C} & \Centerstack{Ref. \citenum{zhang2018hierarchy} \\ Model C*01} & \Centerstack{Ref. \citenum{zhang2018hierarchy} \\ Model C*02} & \Centerstack{Ref. \citenum{zhang2018hierarchy} \\ Model C*03} &\\ \hline
$J_{zz}^{(1)}$     & 0.164        & 0.175              & $\pm$0.02     & 0.149    & 0.149    & 0.149    & 0.149 	&\\
$J_{\pm}^{(1)}$    & 0.108        & 0.108              & $\pm$0.016    & 0.085    & 0.085    & 0.085    & 0.085 	&\\
$J_{\pm\pm}^{(1)}$ & 0            & 0.005              & $\pm$0.019    & 0.07     & 0	     & 0		& 0			&\\
$J_{z\pm}^{(1)}$ 	 & 0.02132      & 0.028              & $\pm$0.014    & 0.1      & 0	     & 0.02132	& -0.02132	&\\
$J_{zz}^{(2)}$     & 0.164        & 0.016              & $\pm$0.012    & 0.02682   & 0.02682 & 0.02682	& 0.02682	&\\
$J_{\pm}^{(2)}$    & 0.0108       & 0.009              & $\pm$0.007    & 0.0153    & 0.0153  & 0.0153   & 0.0153	&\\
$g_{\parallel}$     & 3.72         &    	             & 			    & 3.81     & 3.81	 & 3.81		& 3.81	&\\
$g_{\perp}$  & 3.06         &   	             & 			    & 3.53     & 3.53	 & 3.53		& 3.53	&\\
$T_{f}$      & 0.16 K       &    	             & 			    & 0.6 K     &  	     &  		&  			&\\
$\Theta_{cw}$& 1.8 K        &    	             & 			    & 1.68 K     &  	     &  		&  			&\\
$f = |\Theta_{cw}|/T_f$& 11.25 &    	         & 			    & 2.8     &  	     &  		&  			&\\
$H_c$-calculated& 2.9684 &    	         			& 		 & 2.7535     & 2.3987  & 2.3987 & 2.3987		&\\
$Q_{order}(H_c)$ & $K$ & & & $M$ & $M$ & $M$ & $M$ &\\
  &    &    &     &    &     & 		& 	&\\
\hline
\end{tabular}
\caption{Exchange parameters and $g$-factors, and other relevant quantities found in this work and compared to Ref. \citenum{zhang2018hierarchy} where possible.  Model C, C*01, and C*02 were used to calculate $S(\textbf{Q})$ for Fig. 4}
\end{table*}
Conservatively, our results indicate an intriguing competition of factors.  The spin liquid signatures observed in YbMgGaO$_4$ can arise from high frustration working in concert with thermal fluctuations even at very low temperatures, and these effects are likely enhanced by the inherent chemical disorder of the system.  The diffuse scattering is greatly enhanced by finite temperature, owing to the very high frustration (see supplementary figure 21).  Intriguingly, the scattering intensity at the $M$ points is particularly sensitive to the quenched disorder when compared to the $K$ points (see supplementary fig. 14).  This can likely be understood by the greater degeneracy of possible orientations of stripe order-type phases on the triangular lattice, which generally can have up to 3 orientations. At the same time, the presence of the field-induced crossover also reaffirms quantum mechanical fluctuations as opposed to spin freezing at temperatures as low as 24 mK.   Along with the revised constraints on exchange parameters, these results invite further inquiries into the ground state of YbMgGaO$_4$.  

The central aim of studies of YbMgGaO$_4$ and QSL candidates more generally is to determine the nature of the ground state and to confirm or reject a QSL interpretation.  In this work, we have considered a more general Hamiltonian than the one considered in previous works by allowing the second nearest neighbor interaction, $J^{(2)}$, to be anisotropic. The optimal solution for an isotropic (Heisenberg) second nearest neighbor interaction  $J^{(2)}$ is $J_{\pm}^{(1)}=0.1004(22)$  meV, $J_{\pm\pm}^{(1)}=0.0000(31)$meV, $J_{z\pm}^{(1)}=0.0291(23)$meV, $J_{zz}^{(1)}=0.1611(45)$, $J^{(2)}=0.0100(16)$ (see yellow volume in supplementary figure 15). It is interesting to note that this optimal solution belongs to a spin-liquid phase of the quantum phase diagram reported in Fig. 4 of Ref. \citenum{zhu2018topography}.  Given the high frustration and consequential sensitivity to thermal fluctuations, as well as the presence of quenched disorder, more studies will be needed to disentangle these various contributions to spin liquid features from a possible QSL state.

The effects of disorder on possible QSL states merits further exploration, and a few recent studies provide intriguing comparisons to YbMgGaO$_4$. In stark contrast to YbMgGaO$_4$, where the magnetic ground state defies long range ordering even as the field is raised above the phase crossover (below the polarized state), the closely related triangular antiferromagnet QSL candidate NaYbO$_2$ antiferromagnetically  orders at a critical applied field\cite{bordelon2019field,ranjith2019field}. On the other hand, spin-liquid-like states, as have been suggested for YbMgGaO$_4$\cite{wu2019randomness}, have been proposed as a consequence of cation mixing in the perovskite Sr$_2$Cu(Te$_{0.5}$W$_{0.5}$)O$_6$\cite{mustonen2018spin}. 

Recently, a few studies have likewise used our data to place constraints on the possible parameters \cite{li2020reinvestigation,wu2020exact}.  We note the conclusions compare favorably despite alternative theoretical techniques.


\section{Conclusion}
In summary, our magnetometry and neutron diffraction experiments show a field-induced crossover, which we compare to DMRG and Monte Carlo simulations to place the strictest systematically developed constraints to date on the possible Hamiltonians describing the system.  The presence of this crossover acts as a litmus test for the possible Hamiltonians for the system, and this approach can be extended to other systems where disorder makes the typical means of measuring exchange parameters more challenging.  The range of data used illustrates the importance of establishing sufficient experimental evidence to determine appropriate model parameters in systems with complex Hamiltonians and especially in the presence of quenched disorder.  Furthermore, our results invite further investigation into the competition of high frustration, quenched disorder, and quantum and thermal fluctuations whose significance persists to the lowest investigated temperatures.  Finally, our constraints suggest a proximity to the possible QSL state in the disorder-free limit for the case of isotropic next-nearest neighbor interactions.  

\section{Acknowledgments}
We are grateful to N. Peter Armitage and Alexander L. Chernyshev for their thoughtful comments and discussion. We are grateful to Andrei T. Savici for help in analyzing the neutron scattering data. A portion of this work was performed at the National High Magnetic Field Laboratory, which is supported by the National Science Foundation Cooperative Agreement No. DMR1157490 and DMR-1644779, the State of Florida and the U.S. Department of Energy. A portion of this research used resources at the Spallation Neutron Source, a DOE Office of Science User Facility operated by the Oak Ridge National Laboratory. The computer modeling presented in this paper used resources of the Oak Ridge Leadership Computing Facility, which is supported by the Office of Science of the U.S. Department of Energy under contract no. DE-AC05-00OR22725. We acknowledge the use of the Analytical Instrumentation Facility (AIF) at North Carolina State University, which is supported by the State of North Carolina and the National Science Foundation (award number ECCS-1542015). W.M.S., Z.S., C.M., and S.H. acknowledge support provided by funding from the Powe Junior Faculty Enhancement Award, and William M. Fairbank Chair in Physics at Duke University. A.S. was supported by the DOE Office of Science, Basic Energy Sciences, Scientific User Facilities Division.   C.D.B. was partially funded by the U.S. Department of Energy Office of Basic Energy Sciences. C.D.B. also acknowledges support from the LANL Directed Research and Development program.

\bibliographystyle{apsrev4-1PRX}
\bibliography{3_references_journal_format.bib}

\begin{thebibliography}{69}%
\makeatletter
\providecommand \@ifxundefined [1]{%
 \@ifx{#1\undefined}
}%
\providecommand \@ifnum [1]{%
 \ifnum #1\expandafter \@firstoftwo
 \else \expandafter \@secondoftwo
 \fi
}%
\providecommand \@ifx [1]{%
 \ifx #1\expandafter \@firstoftwo
 \else \expandafter \@secondoftwo
 \fi
}%
\providecommand \natexlab [1]{#1}%
\providecommand \enquote  [1]{``#1''}%
\providecommand \bibnamefont  [1]{#1}%
\providecommand \bibfnamefont [1]{#1}%
\providecommand \citenamefont [1]{#1}%
\providecommand \href@noop [0]{\@secondoftwo}%
\providecommand \href [0]{\begingroup \@sanitize@url \@href}%
\providecommand \@href[1]{\@@startlink{#1}\@@href}%
\providecommand \@@href[1]{\endgroup#1\@@endlink}%
\providecommand \@sanitize@url [0]{\catcode `\\12\catcode `\$12\catcode
  `\&12\catcode `\#12\catcode `\^12\catcode `\_12\catcode `\%12\relax}%
\providecommand \@@startlink[1]{}%
\providecommand \@@endlink[0]{}%
\providecommand \url  [0]{\begingroup\@sanitize@url \@url }%
\providecommand \@url [1]{\endgroup\@href {#1}{\urlprefix }}%
\providecommand \urlprefix  [0]{URL }%
\providecommand \Eprint [0]{\href }%
\providecommand \doibase [0]{http://dx.doi.org/}%
\providecommand \selectlanguage [0]{\@gobble}%
\providecommand \bibinfo  [0]{\@secondoftwo}%
\providecommand \bibfield  [0]{\@secondoftwo}%
\providecommand \translation [1]{[#1]}%
\providecommand \BibitemOpen [0]{}%
\providecommand \bibitemStop [0]{}%
\providecommand \bibitemNoStop [0]{.\EOS\space}%
\providecommand \EOS [0]{\spacefactor3000\relax}%
\providecommand \BibitemShut  [1]{\csname bibitem#1\endcsname}%
\let\auto@bib@innerbib\@empty
\bibitem [{\citenamefont {Savary}\ and\ \citenamefont
  {Balents}(2016)}]{savary2016quantum}%
  \BibitemOpen
  \bibfield  {author} {\bibinfo {author} {\bibfnamefont {L.}~\bibnamefont
  {Savary}}\ and\ \bibinfo {author} {\bibfnamefont {L.}~\bibnamefont
  {Balents}},\ }\bibfield  {title} {\emph {\bibinfo {title} {Quantum spin
  liquids: a review},\ }}\href@noop {} {\bibfield  {journal} {\bibinfo
  {journal} {Rep. Prog. Phys.}\ }\textbf {\bibinfo {volume} {80}},\ \bibinfo
  {pages} {016502} (\bibinfo {year} {2016})}\BibitemShut {NoStop}%
\bibitem [{\citenamefont {Wen}\ \emph {et~al.}(2019)\citenamefont {Wen},
  \citenamefont {Yu}, \citenamefont {Li}, \citenamefont {Yu},\ and\
  \citenamefont {Li}}]{wen2019experimental}%
  \BibitemOpen
  \bibfield  {author} {\bibinfo {author} {\bibfnamefont {J.}~\bibnamefont
  {Wen}}, \bibinfo {author} {\bibfnamefont {S.-L.}\ \bibnamefont {Yu}},
  \bibinfo {author} {\bibfnamefont {S.}~\bibnamefont {Li}}, \bibinfo {author}
  {\bibfnamefont {W.}~\bibnamefont {Yu}}, \ and\ \bibinfo {author}
  {\bibfnamefont {J.-X.}\ \bibnamefont {Li}},\ }\bibfield  {title} {\emph
  {\bibinfo {title} {Experimental identification of quantum spin liquids},\
  }}\href@noop {} {\bibfield  {journal} {\bibinfo  {journal} {npj Quantum
  Mater.}\ }\textbf {\bibinfo {volume} {4}},\ \bibinfo {pages} {1} (\bibinfo
  {year} {2019})}\BibitemShut {NoStop}%
\bibitem [{\citenamefont {Anderson}(1973)}]{anderson1973resonating}%
  \BibitemOpen
  \bibfield  {author} {\bibinfo {author} {\bibfnamefont {P.~W.}\ \bibnamefont
  {Anderson}},\ }\bibfield  {title} {\emph {\bibinfo {title} {Resonating
  valence bonds: A new kind of insulator?}\ }}\href@noop {} {\bibfield
  {journal} {\bibinfo  {journal} {Mat. Res. Bull.}\ }\textbf {\bibinfo {volume}
  {8}},\ \bibinfo {pages} {153} (\bibinfo {year} {1973})}\BibitemShut {NoStop}%
\bibitem [{\citenamefont {Zhong}\ \emph {et~al.}(2019)\citenamefont {Zhong},
  \citenamefont {Guo}, \citenamefont {Xu}, \citenamefont {Xu},\ and\
  \citenamefont {Cava}}]{zhong2019strong}%
  \BibitemOpen
  \bibfield  {author} {\bibinfo {author} {\bibfnamefont {R.}~\bibnamefont
  {Zhong}}, \bibinfo {author} {\bibfnamefont {S.}~\bibnamefont {Guo}}, \bibinfo
  {author} {\bibfnamefont {G.}~\bibnamefont {Xu}}, \bibinfo {author}
  {\bibfnamefont {S.}~\bibnamefont {Xu}}, \ and\ \bibinfo {author}
  {\bibfnamefont {R.}~\bibnamefont {Cava}},\ }\bibfield  {title} {\emph
  {\bibinfo {title} {Strong quantum fluctuations in a quantum spin liquid
  candidate with a {C}o-based triangular lattice},\ }}\href@noop {} {\bibfield
  {journal} {\bibinfo  {journal} {Proc. Natl. Acad. Sci. U.S.A.}\ }\textbf
  {\bibinfo {volume} {116}},\ \bibinfo {pages} {14505–14510} (\bibinfo {year}
  {2019})}\BibitemShut {NoStop}%
\bibitem [{\citenamefont {Bordelon}\ \emph {et~al.}(2019)\citenamefont
  {Bordelon}, \citenamefont {Kenney}, \citenamefont {Liu}, \citenamefont
  {Hogan}, \citenamefont {Posthuma}, \citenamefont {Kavand}, \citenamefont
  {Lyu}, \citenamefont {Sherwin}, \citenamefont {Butch}, \citenamefont {Brown}
  \emph {et~al.}}]{bordelon2019field}%
  \BibitemOpen
  \bibfield  {author} {\bibinfo {author} {\bibfnamefont {M.~M.}\ \bibnamefont
  {Bordelon}}, \bibinfo {author} {\bibfnamefont {E.}~\bibnamefont {Kenney}},
  \bibinfo {author} {\bibfnamefont {C.}~\bibnamefont {Liu}}, \bibinfo {author}
  {\bibfnamefont {T.}~\bibnamefont {Hogan}}, \bibinfo {author} {\bibfnamefont
  {L.}~\bibnamefont {Posthuma}}, \bibinfo {author} {\bibfnamefont
  {M.}~\bibnamefont {Kavand}}, \bibinfo {author} {\bibfnamefont
  {Y.}~\bibnamefont {Lyu}}, \bibinfo {author} {\bibfnamefont {M.}~\bibnamefont
  {Sherwin}}, \bibinfo {author} {\bibfnamefont {N.~P.}\ \bibnamefont {Butch}},
  \bibinfo {author} {\bibfnamefont {C.}~\bibnamefont {Brown}},  \emph
  {et~al.},\ }\bibfield  {title} {\emph {\bibinfo {title} {Field-tunable
  quantum disordered ground state in the triangular-lattice antiferromagnet
  {$NaYbO_2$}},\ }}\href@noop {} {\bibfield  {journal} {\bibinfo  {journal}
  {Nat. Phys.}\ }\textbf {\bibinfo {volume} {15}},\ \bibinfo {pages} {1058}
  (\bibinfo {year} {2019})}\BibitemShut {NoStop}%
\bibitem [{\citenamefont {Liu}\ \emph {et~al.}(2018)\citenamefont {Liu},
  \citenamefont {Zhang}, \citenamefont {Ji}, \citenamefont {Liu}, \citenamefont
  {Li}, \citenamefont {Wang}, \citenamefont {Lei}, \citenamefont {Chen},\ and\
  \citenamefont {Zhang}}]{liu2018rare}%
  \BibitemOpen
  \bibfield  {author} {\bibinfo {author} {\bibfnamefont {W.}~\bibnamefont
  {Liu}}, \bibinfo {author} {\bibfnamefont {Z.}~\bibnamefont {Zhang}}, \bibinfo
  {author} {\bibfnamefont {J.}~\bibnamefont {Ji}}, \bibinfo {author}
  {\bibfnamefont {Y.}~\bibnamefont {Liu}}, \bibinfo {author} {\bibfnamefont
  {J.}~\bibnamefont {Li}}, \bibinfo {author} {\bibfnamefont {X.}~\bibnamefont
  {Wang}}, \bibinfo {author} {\bibfnamefont {H.}~\bibnamefont {Lei}}, \bibinfo
  {author} {\bibfnamefont {G.}~\bibnamefont {Chen}}, \ and\ \bibinfo {author}
  {\bibfnamefont {Q.}~\bibnamefont {Zhang}},\ }\bibfield  {title} {\emph
  {\bibinfo {title} {Rare-earth chalcogenides: A large family of triangular
  lattice spin liquid candidates},\ }}\href@noop {} {\bibfield  {journal}
  {\bibinfo  {journal} {Chin. Phys. Lett.}\ }\textbf {\bibinfo {volume} {35}},\
  \bibinfo {pages} {117501} (\bibinfo {year} {2018})}\BibitemShut {NoStop}%
\bibitem [{\citenamefont {Ashtar}\ \emph {et~al.}(2019)\citenamefont {Ashtar},
  \citenamefont {Marwat}, \citenamefont {Gao}, \citenamefont {Zhang},
  \citenamefont {Pi}, \citenamefont {Yuan},\ and\ \citenamefont
  {Tian}}]{ashtar2019reznal}%
  \BibitemOpen
  \bibfield  {author} {\bibinfo {author} {\bibfnamefont {M.}~\bibnamefont
  {Ashtar}}, \bibinfo {author} {\bibfnamefont {M.}~\bibnamefont {Marwat}},
  \bibinfo {author} {\bibfnamefont {Y.}~\bibnamefont {Gao}}, \bibinfo {author}
  {\bibfnamefont {Z.}~\bibnamefont {Zhang}}, \bibinfo {author} {\bibfnamefont
  {L.}~\bibnamefont {Pi}}, \bibinfo {author} {\bibfnamefont {S.}~\bibnamefont
  {Yuan}}, \ and\ \bibinfo {author} {\bibfnamefont {Z.}~\bibnamefont {Tian}},\
  }\bibfield  {title} {\emph {\bibinfo {title} {{$REZnAl_{11}O_{19}(RE= Pr, Nd,
  Sm-Tb)$}: a new family of ideal 2{D} triangular lattice frustrated magnets},\
  }}\href@noop {} {\bibfield  {journal} {\bibinfo  {journal} {J. Mater. Chem.
  C}\ }\textbf {\bibinfo {volume} {7}},\ \bibinfo {pages} {10073} (\bibinfo
  {year} {2019})}\BibitemShut {NoStop}%
\bibitem [{\citenamefont {Sarkar}\ \emph {et~al.}(2019)\citenamefont {Sarkar},
  \citenamefont {Schlender}, \citenamefont {Grinenko}, \citenamefont
  {Haeussler}, \citenamefont {Baker}, \citenamefont {Doert},\ and\
  \citenamefont {Klauss}}]{sarkar2019quantum}%
  \BibitemOpen
  \bibfield  {author} {\bibinfo {author} {\bibfnamefont {R.}~\bibnamefont
  {Sarkar}}, \bibinfo {author} {\bibfnamefont {P.}~\bibnamefont {Schlender}},
  \bibinfo {author} {\bibfnamefont {V.}~\bibnamefont {Grinenko}}, \bibinfo
  {author} {\bibfnamefont {E.}~\bibnamefont {Haeussler}}, \bibinfo {author}
  {\bibfnamefont {P.~J.}\ \bibnamefont {Baker}}, \bibinfo {author}
  {\bibfnamefont {T.}~\bibnamefont {Doert}}, \ and\ \bibinfo {author}
  {\bibfnamefont {H.-H.}\ \bibnamefont {Klauss}},\ }\bibfield  {title} {\emph
  {\bibinfo {title} {Quantum spin liquid ground state in the disorder free
  triangular lattice {$NaYbS_2$}},\ }}\href@noop {} {\bibfield  {journal}
  {\bibinfo  {journal} {Phys. Rev. B}\ }\textbf {\bibinfo {volume} {100}},\
  \bibinfo {pages} {241116} (\bibinfo {year} {2019})}\BibitemShut {NoStop}%
\bibitem [{\citenamefont {F\aa{}k}\ \emph {et~al.}(2017)\citenamefont
  {F\aa{}k}, \citenamefont {Bieri}, \citenamefont {Can\'evet}, \citenamefont
  {Messio}, \citenamefont {Payen}, \citenamefont {Viaud}, \citenamefont
  {Guillot-Deudon}, \citenamefont {Darie}, \citenamefont {Ollivier},\ and\
  \citenamefont {Mendels}}]{PhysRevB.95.060402}%
  \BibitemOpen
  \bibfield  {author} {\bibinfo {author} {\bibfnamefont {B.}~\bibnamefont
  {F\aa{}k}}, \bibinfo {author} {\bibfnamefont {S.}~\bibnamefont {Bieri}},
  \bibinfo {author} {\bibfnamefont {E.}~\bibnamefont {Can\'evet}}, \bibinfo
  {author} {\bibfnamefont {L.}~\bibnamefont {Messio}}, \bibinfo {author}
  {\bibfnamefont {C.}~\bibnamefont {Payen}}, \bibinfo {author} {\bibfnamefont
  {M.}~\bibnamefont {Viaud}}, \bibinfo {author} {\bibfnamefont
  {C.}~\bibnamefont {Guillot-Deudon}}, \bibinfo {author} {\bibfnamefont
  {C.}~\bibnamefont {Darie}}, \bibinfo {author} {\bibfnamefont
  {J.}~\bibnamefont {Ollivier}}, \ and\ \bibinfo {author} {\bibfnamefont
  {P.}~\bibnamefont {Mendels}},\ }\bibfield  {title} {\emph {\bibinfo {title}
  {Evidence for a spinon {F}ermi surface in the triangular ${S}=1$ quantum spin
  liquid ${Ba}_{3}{N}i{S}b_{2}o_{9}$},\ }}\href {\doibase
  10.1103/PhysRevB.95.060402} {\bibfield  {journal} {\bibinfo  {journal} {Phys.
  Rev. B}\ }\textbf {\bibinfo {volume} {95}},\ \bibinfo {pages} {060402}
  (\bibinfo {year} {2017})}\BibitemShut {NoStop}%
\bibitem [{\citenamefont {Cui}\ \emph {et~al.}(2018)\citenamefont {Cui},
  \citenamefont {Dai}, \citenamefont {Zhou}, \citenamefont {Wang},
  \citenamefont {Li}, \citenamefont {Song}, \citenamefont {Wang}, \citenamefont
  {Ma}, \citenamefont {Zhang}, \citenamefont {Li}, \citenamefont {Luke},
  \citenamefont {Normand}, \citenamefont {Xiang},\ and\ \citenamefont
  {Yu}}]{PhysRevMaterials.2.044403}%
  \BibitemOpen
  \bibfield  {author} {\bibinfo {author} {\bibfnamefont {Y.}~\bibnamefont
  {Cui}}, \bibinfo {author} {\bibfnamefont {J.}~\bibnamefont {Dai}}, \bibinfo
  {author} {\bibfnamefont {P.}~\bibnamefont {Zhou}}, \bibinfo {author}
  {\bibfnamefont {P.~S.}\ \bibnamefont {Wang}}, \bibinfo {author}
  {\bibfnamefont {T.~R.}\ \bibnamefont {Li}}, \bibinfo {author} {\bibfnamefont
  {W.~H.}\ \bibnamefont {Song}}, \bibinfo {author} {\bibfnamefont {J.~C.}\
  \bibnamefont {Wang}}, \bibinfo {author} {\bibfnamefont {L.}~\bibnamefont
  {Ma}}, \bibinfo {author} {\bibfnamefont {Z.}~\bibnamefont {Zhang}}, \bibinfo
  {author} {\bibfnamefont {S.~Y.}\ \bibnamefont {Li}}, \bibinfo {author}
  {\bibfnamefont {G.~M.}\ \bibnamefont {Luke}}, \bibinfo {author}
  {\bibfnamefont {B.}~\bibnamefont {Normand}}, \bibinfo {author} {\bibfnamefont
  {T.}~\bibnamefont {Xiang}}, \ and\ \bibinfo {author} {\bibfnamefont
  {W.}~\bibnamefont {Yu}},\ }\bibfield  {title} {\emph {\bibinfo {title}
  {Mermin-{W}agner physics, {$(H,T)$} phase diagram, and candidate quantum
  spin-liquid phase in the spin-$\frac{1}{2}$ triangular-lattice
  antiferromagnet ${Ba_8CoNb_6O_24}$},\ }}\href {\doibase
  10.1103/PhysRevMaterials.2.044403} {\bibfield  {journal} {\bibinfo  {journal}
  {Phys. Rev. Materials}\ }\textbf {\bibinfo {volume} {2}},\ \bibinfo {pages}
  {044403} (\bibinfo {year} {2018})}\BibitemShut {NoStop}%
\bibitem [{\citenamefont {Iizuka}\ \emph {et~al.}(2020)\citenamefont {Iizuka},
  \citenamefont {Michimura}, \citenamefont {Numakura}, \citenamefont
  {Uwatoko},\ and\ \citenamefont {Kosaka}}]{iizuka2020single}%
  \BibitemOpen
  \bibfield  {author} {\bibinfo {author} {\bibfnamefont {R.}~\bibnamefont
  {Iizuka}}, \bibinfo {author} {\bibfnamefont {S.}~\bibnamefont {Michimura}},
  \bibinfo {author} {\bibfnamefont {R.}~\bibnamefont {Numakura}}, \bibinfo
  {author} {\bibfnamefont {Y.}~\bibnamefont {Uwatoko}}, \ and\ \bibinfo
  {author} {\bibfnamefont {M.}~\bibnamefont {Kosaka}},\ }\bibfield  {title}
  {\emph {\bibinfo {title} {Single crystal growth and physical properties of
  ytterbium sulfide {$KYbS_2$} with triangular lattice},\ }}\href@noop {}
  {\bibfield  {journal} {\bibinfo  {journal} {JPS Conf. Proc.}\ }\textbf
  {\bibinfo {volume} {30}},\ \bibinfo {pages} {011097} (\bibinfo {year}
  {2020})}\BibitemShut {NoStop}%
\bibitem [{\citenamefont {Xing}\ \emph
  {et~al.}(2019{\natexlab{a}})\citenamefont {Xing}, \citenamefont {Sanjeewa},
  \citenamefont {Kim}, \citenamefont {Meier}, \citenamefont {May},
  \citenamefont {Zheng}, \citenamefont {Custelcean}, \citenamefont {Stewart},\
  and\ \citenamefont {Sefat}}]{xing2019synthesis}%
  \BibitemOpen
  \bibfield  {author} {\bibinfo {author} {\bibfnamefont {J.}~\bibnamefont
  {Xing}}, \bibinfo {author} {\bibfnamefont {L.~D.}\ \bibnamefont {Sanjeewa}},
  \bibinfo {author} {\bibfnamefont {J.}~\bibnamefont {Kim}}, \bibinfo {author}
  {\bibfnamefont {W.~R.}\ \bibnamefont {Meier}}, \bibinfo {author}
  {\bibfnamefont {A.~F.}\ \bibnamefont {May}}, \bibinfo {author} {\bibfnamefont
  {Q.}~\bibnamefont {Zheng}}, \bibinfo {author} {\bibfnamefont
  {R.}~\bibnamefont {Custelcean}}, \bibinfo {author} {\bibfnamefont {G.~R.}\
  \bibnamefont {Stewart}}, \ and\ \bibinfo {author} {\bibfnamefont {A.~S.}\
  \bibnamefont {Sefat}},\ }\bibfield  {title} {\emph {\bibinfo {title}
  {Synthesis, magnetization, and heat capacity of triangular lattice materials
  ${NaErSe}_2$ and ${KErSe}_2$},\ }}\href {\doibase
  10.1103/PhysRevMaterials.3.114413} {\bibfield  {journal} {\bibinfo  {journal}
  {Phys. Rev. Mater.}\ }\textbf {\bibinfo {volume} {3}},\ \bibinfo {pages}
  {114413} (\bibinfo {year} {2019}{\natexlab{a}})}\BibitemShut {NoStop}%
\bibitem [{\citenamefont {Gingras}\ and\ \citenamefont
  {McClarty}(2014)}]{gingras2014quantum}%
  \BibitemOpen
  \bibfield  {author} {\bibinfo {author} {\bibfnamefont {M.~J.}\ \bibnamefont
  {Gingras}}\ and\ \bibinfo {author} {\bibfnamefont {P.~A.}\ \bibnamefont
  {McClarty}},\ }\bibfield  {title} {\emph {\bibinfo {title} {Quantum spin ice:
  a search for gapless quantum spin liquids in pyrochlore magnets},\
  }}\href@noop {} {\bibfield  {journal} {\bibinfo  {journal} {Rep. Prog.
  Phys.}\ }\textbf {\bibinfo {volume} {77}},\ \bibinfo {pages} {056501}
  (\bibinfo {year} {2014})}\BibitemShut {NoStop}%
\bibitem [{\citenamefont {Knolle}\ and\ \citenamefont
  {Moessner}(2019)}]{knolle2019field}%
  \BibitemOpen
  \bibfield  {author} {\bibinfo {author} {\bibfnamefont {J.}~\bibnamefont
  {Knolle}}\ and\ \bibinfo {author} {\bibfnamefont {R.}~\bibnamefont
  {Moessner}},\ }\bibfield  {title} {\emph {\bibinfo {title} {A field guide to
  spin liquids},\ }}\href@noop {} {\bibfield  {journal} {\bibinfo  {journal}
  {Annu. Rev. Condens. Matter Phys.}\ }\textbf {\bibinfo {volume} {10}},\
  \bibinfo {pages} {451} (\bibinfo {year} {2019})}\BibitemShut {NoStop}%
\bibitem [{\citenamefont {Suzuki}\ \emph {et~al.}(2020)\citenamefont {Suzuki},
  \citenamefont {Liu}, \citenamefont {Bertinshaw}, \citenamefont {Ueda},
  \citenamefont {Kim}, \citenamefont {Laha}, \citenamefont {Weber},
  \citenamefont {Yang}, \citenamefont {Wang}, \citenamefont {Takahashi},
  \citenamefont {Fürsich}, \citenamefont {Minola}, \citenamefont {Lotsch},
  \citenamefont {Kim}, \citenamefont {Yavaş}, \citenamefont {Daghofer},
  \citenamefont {Chaloupka}, \citenamefont {Khaliullin}, \citenamefont
  {Gretarsson},\ and\ \citenamefont {Keimer}}]{suzuki2020quantifying}%
  \BibitemOpen
  \bibfield  {author} {\bibinfo {author} {\bibfnamefont {H.}~\bibnamefont
  {Suzuki}}, \bibinfo {author} {\bibfnamefont {H.}~\bibnamefont {Liu}},
  \bibinfo {author} {\bibfnamefont {J.}~\bibnamefont {Bertinshaw}}, \bibinfo
  {author} {\bibfnamefont {K.}~\bibnamefont {Ueda}}, \bibinfo {author}
  {\bibfnamefont {H.}~\bibnamefont {Kim}}, \bibinfo {author} {\bibfnamefont
  {S.}~\bibnamefont {Laha}}, \bibinfo {author} {\bibfnamefont {D.}~\bibnamefont
  {Weber}}, \bibinfo {author} {\bibfnamefont {Z.}~\bibnamefont {Yang}},
  \bibinfo {author} {\bibfnamefont {L.}~\bibnamefont {Wang}}, \bibinfo {author}
  {\bibfnamefont {H.}~\bibnamefont {Takahashi}}, \bibinfo {author}
  {\bibfnamefont {K.}~\bibnamefont {Fürsich}}, \bibinfo {author}
  {\bibfnamefont {M.}~\bibnamefont {Minola}}, \bibinfo {author} {\bibfnamefont
  {B.~V.}\ \bibnamefont {Lotsch}}, \bibinfo {author} {\bibfnamefont {B.~J.}\
  \bibnamefont {Kim}}, \bibinfo {author} {\bibfnamefont {H.}~\bibnamefont
  {Yavaş}}, \bibinfo {author} {\bibfnamefont {M.}~\bibnamefont {Daghofer}},
  \bibinfo {author} {\bibfnamefont {J.}~\bibnamefont {Chaloupka}}, \bibinfo
  {author} {\bibfnamefont {G.}~\bibnamefont {Khaliullin}}, \bibinfo {author}
  {\bibfnamefont {H.}~\bibnamefont {Gretarsson}}, \ and\ \bibinfo {author}
  {\bibfnamefont {B.}~\bibnamefont {Keimer}},\ }\bibfield  {title} {\emph
  {\bibinfo {title} {Quantifying the exchange interactions in the {K}itaev
  model system ${R}u{C}l_{3}$ by {R}u ${L}$ edge resonant inelastic x-ray
  scattering},\ }}\href@noop {} {\bibfield  {journal} {\bibinfo  {journal}
  {arXiv preprint arXiv:2008.02037}\ } (\bibinfo {year} {2020})},\ \Eprint
  {http://arxiv.org/abs/2008.02037} {arXiv:2008.02037 [cond-mat.str-el]}
  \BibitemShut {NoStop}%
\bibitem [{\citenamefont {Chillal}\ \emph {et~al.}(2020)\citenamefont
  {Chillal}, \citenamefont {Iqbal}, \citenamefont {Jeschke}, \citenamefont
  {Rodriguez-Rivera}, \citenamefont {Bewley}, \citenamefont {Manuel},
  \citenamefont {Khalyavin}, \citenamefont {Steffens}, \citenamefont {Thomale},
  \citenamefont {Islam} \emph {et~al.}}]{chillal2020evidence}%
  \BibitemOpen
  \bibfield  {author} {\bibinfo {author} {\bibfnamefont {S.}~\bibnamefont
  {Chillal}}, \bibinfo {author} {\bibfnamefont {Y.}~\bibnamefont {Iqbal}},
  \bibinfo {author} {\bibfnamefont {H.~O.}\ \bibnamefont {Jeschke}}, \bibinfo
  {author} {\bibfnamefont {J.~A.}\ \bibnamefont {Rodriguez-Rivera}}, \bibinfo
  {author} {\bibfnamefont {R.}~\bibnamefont {Bewley}}, \bibinfo {author}
  {\bibfnamefont {P.}~\bibnamefont {Manuel}}, \bibinfo {author} {\bibfnamefont
  {D.}~\bibnamefont {Khalyavin}}, \bibinfo {author} {\bibfnamefont
  {P.}~\bibnamefont {Steffens}}, \bibinfo {author} {\bibfnamefont
  {R.}~\bibnamefont {Thomale}}, \bibinfo {author} {\bibfnamefont {A.~N.}\
  \bibnamefont {Islam}},  \emph {et~al.},\ }\bibfield  {title} {\emph {\bibinfo
  {title} {Evidence for a three-dimensional quantum spin liquid in pbcute 2 o
  6},\ }}\href@noop {} {\bibfield  {journal} {\bibinfo  {journal} {Nat.
  Commun.}\ }\textbf {\bibinfo {volume} {11}},\ \bibinfo {pages} {1} (\bibinfo
  {year} {2020})}\BibitemShut {NoStop}%
\bibitem [{\citenamefont {Zhang}\ \emph {et~al.}(2018)\citenamefont {Zhang},
  \citenamefont {Mahmood}, \citenamefont {Daum}, \citenamefont {Dun},
  \citenamefont {Paddison}, \citenamefont {Laurita}, \citenamefont {Hong},
  \citenamefont {Zhou}, \citenamefont {Armitage},\ and\ \citenamefont
  {Mourigal}}]{zhang2018hierarchy}%
  \BibitemOpen
  \bibfield  {author} {\bibinfo {author} {\bibfnamefont {X.}~\bibnamefont
  {Zhang}}, \bibinfo {author} {\bibfnamefont {F.}~\bibnamefont {Mahmood}},
  \bibinfo {author} {\bibfnamefont {M.}~\bibnamefont {Daum}}, \bibinfo {author}
  {\bibfnamefont {Z.}~\bibnamefont {Dun}}, \bibinfo {author} {\bibfnamefont
  {J.~A.}\ \bibnamefont {Paddison}}, \bibinfo {author} {\bibfnamefont {N.~J.}\
  \bibnamefont {Laurita}}, \bibinfo {author} {\bibfnamefont {T.}~\bibnamefont
  {Hong}}, \bibinfo {author} {\bibfnamefont {H.}~\bibnamefont {Zhou}}, \bibinfo
  {author} {\bibfnamefont {N.}~\bibnamefont {Armitage}}, \ and\ \bibinfo
  {author} {\bibfnamefont {M.}~\bibnamefont {Mourigal}},\ }\bibfield  {title}
  {\emph {\bibinfo {title} {Hierarchy of exchange interactions in the
  triangular-lattice spin liquid {$YbMgGaO_4$}},\ }}\href@noop {} {\bibfield
  {journal} {\bibinfo  {journal} {Phys. Rev. X}\ }\textbf {\bibinfo {volume}
  {8}},\ \bibinfo {pages} {031001} (\bibinfo {year} {2018})}\BibitemShut
  {NoStop}%
\bibitem [{\citenamefont {Li}\ \emph {et~al.}(2015{\natexlab{a}})\citenamefont
  {Li}, \citenamefont {Liao}, \citenamefont {Zhang}, \citenamefont {Li},
  \citenamefont {Jin}, \citenamefont {Ling}, \citenamefont {Zhang},
  \citenamefont {Zou}, \citenamefont {Pi}, \citenamefont {Yang} \emph
  {et~al.}}]{li2015gapless}%
  \BibitemOpen
  \bibfield  {author} {\bibinfo {author} {\bibfnamefont {Y.}~\bibnamefont
  {Li}}, \bibinfo {author} {\bibfnamefont {H.}~\bibnamefont {Liao}}, \bibinfo
  {author} {\bibfnamefont {Z.}~\bibnamefont {Zhang}}, \bibinfo {author}
  {\bibfnamefont {S.}~\bibnamefont {Li}}, \bibinfo {author} {\bibfnamefont
  {F.}~\bibnamefont {Jin}}, \bibinfo {author} {\bibfnamefont {L.}~\bibnamefont
  {Ling}}, \bibinfo {author} {\bibfnamefont {L.}~\bibnamefont {Zhang}},
  \bibinfo {author} {\bibfnamefont {Y.}~\bibnamefont {Zou}}, \bibinfo {author}
  {\bibfnamefont {L.}~\bibnamefont {Pi}}, \bibinfo {author} {\bibfnamefont
  {Z.}~\bibnamefont {Yang}},  \emph {et~al.},\ }\bibfield  {title} {\emph
  {\bibinfo {title} {Gapless quantum spin liquid ground state in the
  two-dimensional spin-1/2 triangular antiferromagnet {$YbMgGaO_4$}},\
  }}\href@noop {} {\bibfield  {journal} {\bibinfo  {journal} {Sci. Rep.}\
  }\textbf {\bibinfo {volume} {5}},\ \bibinfo {pages} {1} (\bibinfo {year}
  {2015}{\natexlab{a}})}\BibitemShut {NoStop}%
\bibitem [{\citenamefont {Zhu}\ \emph {et~al.}(2017)\citenamefont {Zhu},
  \citenamefont {Maksimov}, \citenamefont {White},\ and\ \citenamefont
  {Chernyshev}}]{zhu2017disorder}%
  \BibitemOpen
  \bibfield  {author} {\bibinfo {author} {\bibfnamefont {Z.}~\bibnamefont
  {Zhu}}, \bibinfo {author} {\bibfnamefont {P.}~\bibnamefont {Maksimov}},
  \bibinfo {author} {\bibfnamefont {S.~R.}\ \bibnamefont {White}}, \ and\
  \bibinfo {author} {\bibfnamefont {A.}~\bibnamefont {Chernyshev}},\ }\bibfield
   {title} {\emph {\bibinfo {title} {Disorder-induced mimicry of a spin liquid
  in {$YbMgGaO_4$}},\ }}\href@noop {} {\bibfield  {journal} {\bibinfo
  {journal} {Phys. Rev. Lett.}\ }\textbf {\bibinfo {volume} {119}},\ \bibinfo
  {pages} {157201} (\bibinfo {year} {2017})}\BibitemShut {NoStop}%
\bibitem [{\citenamefont {Liu}\ \emph {et~al.}(2016)\citenamefont {Liu},
  \citenamefont {Yu},\ and\ \citenamefont {Wang}}]{liu2016semiclassical}%
  \BibitemOpen
  \bibfield  {author} {\bibinfo {author} {\bibfnamefont {C.}~\bibnamefont
  {Liu}}, \bibinfo {author} {\bibfnamefont {R.}~\bibnamefont {Yu}}, \ and\
  \bibinfo {author} {\bibfnamefont {X.}~\bibnamefont {Wang}},\ }\bibfield
  {title} {\emph {\bibinfo {title} {Semiclassical ground-state phase diagram
  and multi-${Q}$ phase of a spin-orbit-coupled model on triangular lattice},\
  }}\href {\doibase 10.1103/PhysRevB.94.174424} {\bibfield  {journal} {\bibinfo
   {journal} {Phys. Rev. B}\ }\textbf {\bibinfo {volume} {94}},\ \bibinfo
  {pages} {174424} (\bibinfo {year} {2016})}\BibitemShut {NoStop}%
\bibitem [{\citenamefont {Kimchi}\ \emph {et~al.}(2018)\citenamefont {Kimchi},
  \citenamefont {Nahum},\ and\ \citenamefont {Senthil}}]{kimchi2018valence}%
  \BibitemOpen
  \bibfield  {author} {\bibinfo {author} {\bibfnamefont {I.}~\bibnamefont
  {Kimchi}}, \bibinfo {author} {\bibfnamefont {A.}~\bibnamefont {Nahum}}, \
  and\ \bibinfo {author} {\bibfnamefont {T.}~\bibnamefont {Senthil}},\
  }\bibfield  {title} {\emph {\bibinfo {title} {Valence bonds in random quantum
  magnets: theory and application to {$YbMgGaO_4$}},\ }}\href@noop {}
  {\bibfield  {journal} {\bibinfo  {journal} {Phys. Rev. X}\ }\textbf {\bibinfo
  {volume} {8}},\ \bibinfo {pages} {031028} (\bibinfo {year}
  {2018})}\BibitemShut {NoStop}%
\bibitem [{\citenamefont {Wu}\ \emph {et~al.}(2019)\citenamefont {Wu},
  \citenamefont {Gong},\ and\ \citenamefont {Sheng}}]{wu2019randomness}%
  \BibitemOpen
  \bibfield  {author} {\bibinfo {author} {\bibfnamefont {H.-Q.}\ \bibnamefont
  {Wu}}, \bibinfo {author} {\bibfnamefont {S.-S.}\ \bibnamefont {Gong}}, \ and\
  \bibinfo {author} {\bibfnamefont {D.}~\bibnamefont {Sheng}},\ }\bibfield
  {title} {\emph {\bibinfo {title} {Randomness-induced spin-liquid-like phase
  in the spin-$1/2$ {$J_1-J_2$} triangular {H}eisenberg model},\ }}\href@noop
  {} {\bibfield  {journal} {\bibinfo  {journal} {Phys. Rev. B}\ }\textbf
  {\bibinfo {volume} {99}},\ \bibinfo {pages} {085141} (\bibinfo {year}
  {2019})}\BibitemShut {NoStop}%
\bibitem [{\citenamefont {Zhu}\ \emph {et~al.}(2018)\citenamefont {Zhu},
  \citenamefont {Maksimov}, \citenamefont {White},\ and\ \citenamefont
  {Chernyshev}}]{zhu2018topography}%
  \BibitemOpen
  \bibfield  {author} {\bibinfo {author} {\bibfnamefont {Z.}~\bibnamefont
  {Zhu}}, \bibinfo {author} {\bibfnamefont {P.}~\bibnamefont {Maksimov}},
  \bibinfo {author} {\bibfnamefont {S.~R.}\ \bibnamefont {White}}, \ and\
  \bibinfo {author} {\bibfnamefont {A.}~\bibnamefont {Chernyshev}},\ }\bibfield
   {title} {\emph {\bibinfo {title} {Topography of spin liquids on a triangular
  lattice},\ }}\href@noop {} {\bibfield  {journal} {\bibinfo  {journal} {Phys.
  Rev. Lett.}\ }\textbf {\bibinfo {volume} {120}},\ \bibinfo {pages} {207203}
  (\bibinfo {year} {2018})}\BibitemShut {NoStop}%
\bibitem [{\citenamefont {Parker}\ and\ \citenamefont
  {Balents}(2018)}]{parker2018finite}%
  \BibitemOpen
  \bibfield  {author} {\bibinfo {author} {\bibfnamefont {E.}~\bibnamefont
  {Parker}}\ and\ \bibinfo {author} {\bibfnamefont {L.}~\bibnamefont
  {Balents}},\ }\bibfield  {title} {\emph {\bibinfo {title} {Finite-temperature
  behavior of a classical spin-orbit-coupled model for {$YbMgGaO_4$} with and
  without bond disorder},\ }}\href@noop {} {\bibfield  {journal} {\bibinfo
  {journal} {arXiv preprint arXiv:1801.06941}\ } (\bibinfo {year}
  {2018})}\BibitemShut {NoStop}%
\bibitem [{\citenamefont {Li}\ \emph {et~al.}(2018)\citenamefont {Li},
  \citenamefont {Shen}, \citenamefont {Li}, \citenamefont {Zhao},\ and\
  \citenamefont {Chen}}]{li2018effect}%
  \BibitemOpen
  \bibfield  {author} {\bibinfo {author} {\bibfnamefont {Y.-D.}\ \bibnamefont
  {Li}}, \bibinfo {author} {\bibfnamefont {Y.}~\bibnamefont {Shen}}, \bibinfo
  {author} {\bibfnamefont {Y.}~\bibnamefont {Li}}, \bibinfo {author}
  {\bibfnamefont {J.}~\bibnamefont {Zhao}}, \ and\ \bibinfo {author}
  {\bibfnamefont {G.}~\bibnamefont {Chen}},\ }\bibfield  {title} {\emph
  {\bibinfo {title} {Effect of spin-orbit coupling on the effective-spin
  correlation in {$YbMgGaO_4$}},\ }}\href@noop {} {\bibfield  {journal}
  {\bibinfo  {journal} {Phys. Rev. B}\ }\textbf {\bibinfo {volume} {97}},\
  \bibinfo {pages} {125105} (\bibinfo {year} {2018})}\BibitemShut {NoStop}%
\bibitem [{\citenamefont {Luo}\ \emph {et~al.}(2018)\citenamefont {Luo},
  \citenamefont {Lake}, \citenamefont {Mei},\ and\ \citenamefont
  {Starykh}}]{luo2018spinon}%
  \BibitemOpen
  \bibfield  {author} {\bibinfo {author} {\bibfnamefont {Z.-X.}\ \bibnamefont
  {Luo}}, \bibinfo {author} {\bibfnamefont {E.}~\bibnamefont {Lake}}, \bibinfo
  {author} {\bibfnamefont {J.-W.}\ \bibnamefont {Mei}}, \ and\ \bibinfo
  {author} {\bibfnamefont {O.~A.}\ \bibnamefont {Starykh}},\ }\bibfield
  {title} {\emph {\bibinfo {title} {Spinon magnetic resonance of quantum spin
  liquids},\ }}\href@noop {} {\bibfield  {journal} {\bibinfo  {journal} {Phys.
  Rev. Lett.}\ }\textbf {\bibinfo {volume} {120}},\ \bibinfo {pages} {037204}
  (\bibinfo {year} {2018})}\BibitemShut {NoStop}%
\bibitem [{\citenamefont {Iaconis}\ \emph {et~al.}(2018)\citenamefont
  {Iaconis}, \citenamefont {Liu}, \citenamefont {Hal{\'a}sz},\ and\
  \citenamefont {Balents}}]{iaconis2018spin}%
  \BibitemOpen
  \bibfield  {author} {\bibinfo {author} {\bibfnamefont {J.}~\bibnamefont
  {Iaconis}}, \bibinfo {author} {\bibfnamefont {C.}~\bibnamefont {Liu}},
  \bibinfo {author} {\bibfnamefont {G.~B.}\ \bibnamefont {Hal{\'a}sz}}, \ and\
  \bibinfo {author} {\bibfnamefont {L.}~\bibnamefont {Balents}},\ }\bibfield
  {title} {\emph {\bibinfo {title} {Spin liquid versus spin orbit coupling on
  the triangular lattice},\ }}\href@noop {} {\bibfield  {journal} {\bibinfo
  {journal} {SciPost Phys.}\ }\textbf {\bibinfo {volume} {4}} (\bibinfo {year}
  {2018})}\BibitemShut {NoStop}%
\bibitem [{\citenamefont {Li}\ \emph {et~al.}(2017{\natexlab{a}})\citenamefont
  {Li}, \citenamefont {Lu},\ and\ \citenamefont {Chen}}]{li2017spinon}%
  \BibitemOpen
  \bibfield  {author} {\bibinfo {author} {\bibfnamefont {Y.-D.}\ \bibnamefont
  {Li}}, \bibinfo {author} {\bibfnamefont {Y.-M.}\ \bibnamefont {Lu}}, \ and\
  \bibinfo {author} {\bibfnamefont {G.}~\bibnamefont {Chen}},\ }\bibfield
  {title} {\emph {\bibinfo {title} {Spinon {F}ermi surface {U}(1) spin liquid
  in the spin-orbit-coupled triangular-lattice {M}ott insulator
  {$YbMgGaO_4$}},\ }}\href@noop {} {\bibfield  {journal} {\bibinfo  {journal}
  {Phys. Rev. B}\ }\textbf {\bibinfo {volume} {96}},\ \bibinfo {pages} {054445}
  (\bibinfo {year} {2017}{\natexlab{a}})}\BibitemShut {NoStop}%
\bibitem [{\citenamefont {Maksimov}\ \emph {et~al.}(2019)\citenamefont
  {Maksimov}, \citenamefont {Zhu}, \citenamefont {White},\ and\ \citenamefont
  {Chernyshev}}]{maksimov2019anisotropic}%
  \BibitemOpen
  \bibfield  {author} {\bibinfo {author} {\bibfnamefont {P.}~\bibnamefont
  {Maksimov}}, \bibinfo {author} {\bibfnamefont {Z.}~\bibnamefont {Zhu}},
  \bibinfo {author} {\bibfnamefont {S.~R.}\ \bibnamefont {White}}, \ and\
  \bibinfo {author} {\bibfnamefont {A.}~\bibnamefont {Chernyshev}},\ }\bibfield
   {title} {\emph {\bibinfo {title} {Anisotropic-exchange magnets on a
  triangular lattice: spin waves, accidental degeneracies, and dual spin
  liquids},\ }}\href@noop {} {\bibfield  {journal} {\bibinfo  {journal} {Phys.
  Rev. X}\ }\textbf {\bibinfo {volume} {9}},\ \bibinfo {pages} {021017}
  (\bibinfo {year} {2019})}\BibitemShut {NoStop}%
\bibitem [{\citenamefont {Lima}(2019)}]{lima2019spatial}%
  \BibitemOpen
  \bibfield  {author} {\bibinfo {author} {\bibfnamefont {M.~P.}\ \bibnamefont
  {Lima}},\ }\bibfield  {title} {\emph {\bibinfo {title} {Spatial anisotropy of
  the quantum spin liquid system {$YbMgGaO_4$} revealed by ab initio
  calculations},\ }}\href@noop {} {\bibfield  {journal} {\bibinfo  {journal}
  {J. Condens. Matter Phys.}\ }\textbf {\bibinfo {volume} {32}},\ \bibinfo
  {pages} {025505} (\bibinfo {year} {2019})}\BibitemShut {NoStop}%
\bibitem [{\citenamefont {Li}\ and\ \citenamefont
  {Chen}(2017)}]{li2017detecting}%
  \BibitemOpen
  \bibfield  {author} {\bibinfo {author} {\bibfnamefont {Y.-D.}\ \bibnamefont
  {Li}}\ and\ \bibinfo {author} {\bibfnamefont {G.}~\bibnamefont {Chen}},\
  }\bibfield  {title} {\emph {\bibinfo {title} {Detecting spin
  fractionalization in a spinon {F}ermi surface spin liquid},\ }}\href@noop {}
  {\bibfield  {journal} {\bibinfo  {journal} {Phys. Rev. B}\ }\textbf {\bibinfo
  {volume} {96}},\ \bibinfo {pages} {075105} (\bibinfo {year}
  {2017})}\BibitemShut {NoStop}%
\bibitem [{\citenamefont {Gong}\ \emph {et~al.}(2017)\citenamefont {Gong},
  \citenamefont {Zhu}, \citenamefont {Zhu}, \citenamefont {Sheng},\ and\
  \citenamefont {Yang}}]{gong2017global}%
  \BibitemOpen
  \bibfield  {author} {\bibinfo {author} {\bibfnamefont {S.-S.}\ \bibnamefont
  {Gong}}, \bibinfo {author} {\bibfnamefont {W.}~\bibnamefont {Zhu}}, \bibinfo
  {author} {\bibfnamefont {J.-X.}\ \bibnamefont {Zhu}}, \bibinfo {author}
  {\bibfnamefont {D.~N.}\ \bibnamefont {Sheng}}, \ and\ \bibinfo {author}
  {\bibfnamefont {K.}~\bibnamefont {Yang}},\ }\bibfield  {title} {\emph
  {\bibinfo {title} {Global phase diagram and quantum spin liquids in a
  spin-$1/2$ triangular antiferromagnet},\ }}\href@noop {} {\bibfield
  {journal} {\bibinfo  {journal} {Phys. Rev. B}\ }\textbf {\bibinfo {volume}
  {96}},\ \bibinfo {pages} {075116} (\bibinfo {year} {2017})}\BibitemShut
  {NoStop}%
\bibitem [{\citenamefont {Luo}\ \emph {et~al.}(2017)\citenamefont {Luo},
  \citenamefont {Hu}, \citenamefont {Xi}, \citenamefont {Zhao},\ and\
  \citenamefont {Wang}}]{luo2017ground}%
  \BibitemOpen
  \bibfield  {author} {\bibinfo {author} {\bibfnamefont {Q.}~\bibnamefont
  {Luo}}, \bibinfo {author} {\bibfnamefont {S.}~\bibnamefont {Hu}}, \bibinfo
  {author} {\bibfnamefont {B.}~\bibnamefont {Xi}}, \bibinfo {author}
  {\bibfnamefont {J.}~\bibnamefont {Zhao}}, \ and\ \bibinfo {author}
  {\bibfnamefont {X.}~\bibnamefont {Wang}},\ }\bibfield  {title} {\emph
  {\bibinfo {title} {Ground-state phase diagram of an anisotropic spin-1/2
  model on the triangular lattice},\ }}\href@noop {} {\bibfield  {journal}
  {\bibinfo  {journal} {Phys. Rev. B}\ }\textbf {\bibinfo {volume} {95}},\
  \bibinfo {pages} {165110} (\bibinfo {year} {2017})}\BibitemShut {NoStop}%
\bibitem [{\citenamefont {Li}(2020)}]{li2020reinvestigation}%
  \BibitemOpen
  \bibfield  {author} {\bibinfo {author} {\bibfnamefont {S.}~\bibnamefont
  {Li}},\ }\bibfield  {title} {\emph {\bibinfo {title} {Reinvestigation of the
  homogeneous spin model in {$YbMgGaO_4$}},\ }}\href@noop {} {\bibfield
  {journal} {\bibinfo  {journal} {arXiv preprint arXiv:2008.06544}\ } (\bibinfo
  {year} {2020})}\BibitemShut {NoStop}%
\bibitem [{\citenamefont {Wu}\ \emph {et~al.}(2020)\citenamefont {Wu},
  \citenamefont {Yao},\ and\ \citenamefont {Wu}}]{wu2020exact}%
  \BibitemOpen
  \bibfield  {author} {\bibinfo {author} {\bibfnamefont {M.}~\bibnamefont
  {Wu}}, \bibinfo {author} {\bibfnamefont {D.-X.}\ \bibnamefont {Yao}}, \ and\
  \bibinfo {author} {\bibfnamefont {H.-Q.}\ \bibnamefont {Wu}},\ }\bibfield
  {title} {\emph {\bibinfo {title} {Exact diagonalization study of the
  anisotropic {H}eisenberg model related to {$YbMgGaO_4$} and $naybch_2$},\
  }}\href@noop {} {\bibfield  {journal} {\bibinfo  {journal} {arXiv preprint
  arXiv:2008.08751}\ } (\bibinfo {year} {2020})}\BibitemShut {NoStop}%
\bibitem [{\citenamefont {Ma}\ \emph {et~al.}(2018{\natexlab{a}})\citenamefont
  {Ma}, \citenamefont {Wang}, \citenamefont {Dong}, \citenamefont {Zhang},
  \citenamefont {Li}, \citenamefont {Zheng}, \citenamefont {Yu}, \citenamefont
  {Wang}, \citenamefont {Che}, \citenamefont {Ran} \emph
  {et~al.}}]{ma2018spin}%
  \BibitemOpen
  \bibfield  {author} {\bibinfo {author} {\bibfnamefont {Z.}~\bibnamefont
  {Ma}}, \bibinfo {author} {\bibfnamefont {J.}~\bibnamefont {Wang}}, \bibinfo
  {author} {\bibfnamefont {Z.-Y.}\ \bibnamefont {Dong}}, \bibinfo {author}
  {\bibfnamefont {J.}~\bibnamefont {Zhang}}, \bibinfo {author} {\bibfnamefont
  {S.}~\bibnamefont {Li}}, \bibinfo {author} {\bibfnamefont {S.-H.}\
  \bibnamefont {Zheng}}, \bibinfo {author} {\bibfnamefont {Y.}~\bibnamefont
  {Yu}}, \bibinfo {author} {\bibfnamefont {W.}~\bibnamefont {Wang}}, \bibinfo
  {author} {\bibfnamefont {L.}~\bibnamefont {Che}}, \bibinfo {author}
  {\bibfnamefont {K.}~\bibnamefont {Ran}},  \emph {et~al.},\ }\bibfield
  {title} {\emph {\bibinfo {title} {Spin-glass ground state in a
  triangular-lattice compound {$YbZnGaO_4$}},\ }}\href@noop {} {\bibfield
  {journal} {\bibinfo  {journal} {Phys. Rev. Lett.}\ }\textbf {\bibinfo
  {volume} {120}},\ \bibinfo {pages} {087201} (\bibinfo {year}
  {2018}{\natexlab{a}})}\BibitemShut {NoStop}%
\bibitem [{\citenamefont {Li}\ \emph {et~al.}(2019)\citenamefont {Li},
  \citenamefont {Bachus}, \citenamefont {Liu}, \citenamefont {Radelytskyi},
  \citenamefont {Bertin}, \citenamefont {Schneidewind}, \citenamefont {Tokiwa},
  \citenamefont {Tsirlin},\ and\ \citenamefont
  {Gegenwart}}]{li2019rearrangement}%
  \BibitemOpen
  \bibfield  {author} {\bibinfo {author} {\bibfnamefont {Y.}~\bibnamefont
  {Li}}, \bibinfo {author} {\bibfnamefont {S.}~\bibnamefont {Bachus}}, \bibinfo
  {author} {\bibfnamefont {B.}~\bibnamefont {Liu}}, \bibinfo {author}
  {\bibfnamefont {I.}~\bibnamefont {Radelytskyi}}, \bibinfo {author}
  {\bibfnamefont {A.}~\bibnamefont {Bertin}}, \bibinfo {author} {\bibfnamefont
  {A.}~\bibnamefont {Schneidewind}}, \bibinfo {author} {\bibfnamefont
  {Y.}~\bibnamefont {Tokiwa}}, \bibinfo {author} {\bibfnamefont {A.~A.}\
  \bibnamefont {Tsirlin}}, \ and\ \bibinfo {author} {\bibfnamefont
  {P.}~\bibnamefont {Gegenwart}},\ }\bibfield  {title} {\emph {\bibinfo {title}
  {Rearrangement of uncorrelated valence bonds evidenced by low-energy spin
  excitations in {$YbMgGaO_4$}},\ }}\href@noop {} {\bibfield  {journal}
  {\bibinfo  {journal} {Phys. Rev. Lett.}\ }\textbf {\bibinfo {volume} {122}},\
  \bibinfo {pages} {137201} (\bibinfo {year} {2019})}\BibitemShut {NoStop}%
\bibitem [{\citenamefont {T{\'o}th}\ \emph {et~al.}(2017)\citenamefont
  {T{\'o}th}, \citenamefont {Rolfs}, \citenamefont {Wildes},\ and\
  \citenamefont {R{\"u}egg}}]{toth2017strong}%
  \BibitemOpen
  \bibfield  {author} {\bibinfo {author} {\bibfnamefont {S.}~\bibnamefont
  {T{\'o}th}}, \bibinfo {author} {\bibfnamefont {K.}~\bibnamefont {Rolfs}},
  \bibinfo {author} {\bibfnamefont {A.~R.}\ \bibnamefont {Wildes}}, \ and\
  \bibinfo {author} {\bibfnamefont {C.}~\bibnamefont {R{\"u}egg}},\ }\bibfield
  {title} {\emph {\bibinfo {title} {Strong exchange anisotropy in {$YbMgGaO_4$}
  from polarized neutron diffraction},\ }}\href@noop {} {\bibfield  {journal}
  {\bibinfo  {journal} {arXiv preprint arXiv:1705.05699}\ } (\bibinfo {year}
  {2017})}\BibitemShut {NoStop}%
\bibitem [{\citenamefont {Li}\ \emph {et~al.}(2015{\natexlab{b}})\citenamefont
  {Li}, \citenamefont {Chen}, \citenamefont {Tong}, \citenamefont {Pi},
  \citenamefont {Liu}, \citenamefont {Yang}, \citenamefont {Wang},\ and\
  \citenamefont {Zhang}}]{li2015rare}%
  \BibitemOpen
  \bibfield  {author} {\bibinfo {author} {\bibfnamefont {Y.}~\bibnamefont
  {Li}}, \bibinfo {author} {\bibfnamefont {G.}~\bibnamefont {Chen}}, \bibinfo
  {author} {\bibfnamefont {W.}~\bibnamefont {Tong}}, \bibinfo {author}
  {\bibfnamefont {L.}~\bibnamefont {Pi}}, \bibinfo {author} {\bibfnamefont
  {J.}~\bibnamefont {Liu}}, \bibinfo {author} {\bibfnamefont {Z.}~\bibnamefont
  {Yang}}, \bibinfo {author} {\bibfnamefont {X.}~\bibnamefont {Wang}}, \ and\
  \bibinfo {author} {\bibfnamefont {Q.}~\bibnamefont {Zhang}},\ }\bibfield
  {title} {\emph {\bibinfo {title} {Rare-earth triangular lattice spin liquid:
  a single-crystal study of {$YbMgGaO_4$}},\ }}\href@noop {} {\bibfield
  {journal} {\bibinfo  {journal} {Phys. Rev. Lett.}\ }\textbf {\bibinfo
  {volume} {115}},\ \bibinfo {pages} {167203} (\bibinfo {year}
  {2015}{\natexlab{b}})}\BibitemShut {NoStop}%
\bibitem [{\citenamefont {Xu}\ \emph {et~al.}(2016)\citenamefont {Xu},
  \citenamefont {Zhang}, \citenamefont {Li}, \citenamefont {Yu}, \citenamefont
  {Hong}, \citenamefont {Zhang},\ and\ \citenamefont {Li}}]{xu2016absence}%
  \BibitemOpen
  \bibfield  {author} {\bibinfo {author} {\bibfnamefont {Y.}~\bibnamefont
  {Xu}}, \bibinfo {author} {\bibfnamefont {J.}~\bibnamefont {Zhang}}, \bibinfo
  {author} {\bibfnamefont {Y.}~\bibnamefont {Li}}, \bibinfo {author}
  {\bibfnamefont {Y.}~\bibnamefont {Yu}}, \bibinfo {author} {\bibfnamefont
  {X.}~\bibnamefont {Hong}}, \bibinfo {author} {\bibfnamefont {Q.}~\bibnamefont
  {Zhang}}, \ and\ \bibinfo {author} {\bibfnamefont {S.}~\bibnamefont {Li}},\
  }\bibfield  {title} {\emph {\bibinfo {title} {Absence of magnetic thermal
  conductivity in the quantum spin-liquid candidate {$YbMgGaO_4$}},\
  }}\href@noop {} {\bibfield  {journal} {\bibinfo  {journal} {Phys. Rev.
  Lett.}\ }\textbf {\bibinfo {volume} {117}},\ \bibinfo {pages} {267202}
  (\bibinfo {year} {2016})}\BibitemShut {NoStop}%
\bibitem [{\citenamefont {Li}\ \emph {et~al.}(2017{\natexlab{b}})\citenamefont
  {Li}, \citenamefont {Adroja}, \citenamefont {Voneshen}, \citenamefont
  {Bewley}, \citenamefont {Zhang}, \citenamefont {Tsirlin},\ and\ \citenamefont
  {Gegenwart}}]{li2017nearest}%
  \BibitemOpen
  \bibfield  {author} {\bibinfo {author} {\bibfnamefont {Y.}~\bibnamefont
  {Li}}, \bibinfo {author} {\bibfnamefont {D.}~\bibnamefont {Adroja}}, \bibinfo
  {author} {\bibfnamefont {D.}~\bibnamefont {Voneshen}}, \bibinfo {author}
  {\bibfnamefont {R.~I.}\ \bibnamefont {Bewley}}, \bibinfo {author}
  {\bibfnamefont {Q.}~\bibnamefont {Zhang}}, \bibinfo {author} {\bibfnamefont
  {A.~A.}\ \bibnamefont {Tsirlin}}, \ and\ \bibinfo {author} {\bibfnamefont
  {P.}~\bibnamefont {Gegenwart}},\ }\bibfield  {title} {\emph {\bibinfo {title}
  {Nearest-neighbour resonating valence bonds in {$YbMgGaO_4$}},\ }}\href@noop
  {} {\bibfield  {journal} {\bibinfo  {journal} {Nat. Commun.}\ }\textbf
  {\bibinfo {volume} {8}},\ \bibinfo {pages} {1} (\bibinfo {year}
  {2017}{\natexlab{b}})}\BibitemShut {NoStop}%
\bibitem [{\citenamefont {Li}\ \emph {et~al.}(2017{\natexlab{c}})\citenamefont
  {Li}, \citenamefont {Adroja}, \citenamefont {Bewley}, \citenamefont
  {Voneshen}, \citenamefont {Tsirlin}, \citenamefont {Gegenwart},\ and\
  \citenamefont {Zhang}}]{li2017crystalline}%
  \BibitemOpen
  \bibfield  {author} {\bibinfo {author} {\bibfnamefont {Y.}~\bibnamefont
  {Li}}, \bibinfo {author} {\bibfnamefont {D.}~\bibnamefont {Adroja}}, \bibinfo
  {author} {\bibfnamefont {R.~I.}\ \bibnamefont {Bewley}}, \bibinfo {author}
  {\bibfnamefont {D.}~\bibnamefont {Voneshen}}, \bibinfo {author}
  {\bibfnamefont {A.~A.}\ \bibnamefont {Tsirlin}}, \bibinfo {author}
  {\bibfnamefont {P.}~\bibnamefont {Gegenwart}}, \ and\ \bibinfo {author}
  {\bibfnamefont {Q.}~\bibnamefont {Zhang}},\ }\bibfield  {title} {\emph
  {\bibinfo {title} {Crystalline electric-field randomness in the triangular
  lattice spin-liquid {$YbMgGaO_4$}},\ }}\href@noop {} {\bibfield  {journal}
  {\bibinfo  {journal} {Phys. Rev. Lett.}\ }\textbf {\bibinfo {volume} {118}},\
  \bibinfo {pages} {107202} (\bibinfo {year} {2017}{\natexlab{c}})}\BibitemShut
  {NoStop}%
\bibitem [{\citenamefont {Paddison}\ \emph {et~al.}(2017)\citenamefont
  {Paddison}, \citenamefont {Daum}, \citenamefont {Dun}, \citenamefont
  {Ehlers}, \citenamefont {Liu}, \citenamefont {Stone}, \citenamefont {Zhou},\
  and\ \citenamefont {Mourigal}}]{paddison2017continuous}%
  \BibitemOpen
  \bibfield  {author} {\bibinfo {author} {\bibfnamefont {J.~A.}\ \bibnamefont
  {Paddison}}, \bibinfo {author} {\bibfnamefont {M.}~\bibnamefont {Daum}},
  \bibinfo {author} {\bibfnamefont {Z.}~\bibnamefont {Dun}}, \bibinfo {author}
  {\bibfnamefont {G.}~\bibnamefont {Ehlers}}, \bibinfo {author} {\bibfnamefont
  {Y.}~\bibnamefont {Liu}}, \bibinfo {author} {\bibfnamefont {M.~B.}\
  \bibnamefont {Stone}}, \bibinfo {author} {\bibfnamefont {H.}~\bibnamefont
  {Zhou}}, \ and\ \bibinfo {author} {\bibfnamefont {M.}~\bibnamefont
  {Mourigal}},\ }\bibfield  {title} {\emph {\bibinfo {title} {Continuous
  excitations of the triangular-lattice quantum spin liquid {$YbMgGaO_4$}},\
  }}\href@noop {} {\bibfield  {journal} {\bibinfo  {journal} {Nat. Phys.}\
  }\textbf {\bibinfo {volume} {13}},\ \bibinfo {pages} {117} (\bibinfo {year}
  {2017})}\BibitemShut {NoStop}%
\bibitem [{\citenamefont {Shen}\ \emph {et~al.}(2018)\citenamefont {Shen},
  \citenamefont {Li}, \citenamefont {Walker}, \citenamefont {Steffens},
  \citenamefont {Boehm}, \citenamefont {Zhang}, \citenamefont {Shen},
  \citenamefont {Wo}, \citenamefont {Chen},\ and\ \citenamefont
  {Zhao}}]{shen2018fractionalized}%
  \BibitemOpen
  \bibfield  {author} {\bibinfo {author} {\bibfnamefont {Y.}~\bibnamefont
  {Shen}}, \bibinfo {author} {\bibfnamefont {Y.-D.}\ \bibnamefont {Li}},
  \bibinfo {author} {\bibfnamefont {H.}~\bibnamefont {Walker}}, \bibinfo
  {author} {\bibfnamefont {P.}~\bibnamefont {Steffens}}, \bibinfo {author}
  {\bibfnamefont {M.}~\bibnamefont {Boehm}}, \bibinfo {author} {\bibfnamefont
  {X.}~\bibnamefont {Zhang}}, \bibinfo {author} {\bibfnamefont
  {S.}~\bibnamefont {Shen}}, \bibinfo {author} {\bibfnamefont {H.}~\bibnamefont
  {Wo}}, \bibinfo {author} {\bibfnamefont {G.}~\bibnamefont {Chen}}, \ and\
  \bibinfo {author} {\bibfnamefont {J.}~\bibnamefont {Zhao}},\ }\bibfield
  {title} {\emph {\bibinfo {title} {Fractionalized excitations in the partially
  magnetized spin liquid candidate {$YbMgGaO_4$}},\ }}\href@noop {} {\bibfield
  {journal} {\bibinfo  {journal} {Nat. Commun.}\ }\textbf {\bibinfo {volume}
  {9}},\ \bibinfo {pages} {1} (\bibinfo {year} {2018})}\BibitemShut {NoStop}%
\bibitem [{\citenamefont {Li}\ \emph {et~al.}(2016{\natexlab{a}})\citenamefont
  {Li}, \citenamefont {Adroja}, \citenamefont {Biswas}, \citenamefont {Baker},
  \citenamefont {Zhang}, \citenamefont {Liu}, \citenamefont {Tsirlin},
  \citenamefont {Gegenwart},\ and\ \citenamefont {Zhang}}]{li2016muon}%
  \BibitemOpen
  \bibfield  {author} {\bibinfo {author} {\bibfnamefont {Y.}~\bibnamefont
  {Li}}, \bibinfo {author} {\bibfnamefont {D.}~\bibnamefont {Adroja}}, \bibinfo
  {author} {\bibfnamefont {P.~K.}\ \bibnamefont {Biswas}}, \bibinfo {author}
  {\bibfnamefont {P.~J.}\ \bibnamefont {Baker}}, \bibinfo {author}
  {\bibfnamefont {Q.}~\bibnamefont {Zhang}}, \bibinfo {author} {\bibfnamefont
  {J.}~\bibnamefont {Liu}}, \bibinfo {author} {\bibfnamefont {A.~A.}\
  \bibnamefont {Tsirlin}}, \bibinfo {author} {\bibfnamefont {P.}~\bibnamefont
  {Gegenwart}}, \ and\ \bibinfo {author} {\bibfnamefont {Q.}~\bibnamefont
  {Zhang}},\ }\bibfield  {title} {\emph {\bibinfo {title} {Muon spin relaxation
  evidence for the {U}(1) quantum spin-liquid ground state in the triangular
  antiferromagnet {$YbMgGaO_4$}},\ }}\href@noop {} {\bibfield  {journal}
  {\bibinfo  {journal} {Phys. Rev. Lett.}\ }\textbf {\bibinfo {volume} {117}},\
  \bibinfo {pages} {097201} (\bibinfo {year} {2016}{\natexlab{a}})}\BibitemShut
  {NoStop}%
\bibitem [{\citenamefont {Bachus}\ \emph {et~al.}(2020)\citenamefont {Bachus},
  \citenamefont {Iakovlev}, \citenamefont {Li}, \citenamefont {W{\"o}rl},
  \citenamefont {Tokiwa}, \citenamefont {Ling}, \citenamefont {Zhang},
  \citenamefont {Mazurenko}, \citenamefont {Gegenwart},\ and\ \citenamefont
  {Tsirlin}}]{bachus2020field}%
  \BibitemOpen
  \bibfield  {author} {\bibinfo {author} {\bibfnamefont {S.}~\bibnamefont
  {Bachus}}, \bibinfo {author} {\bibfnamefont {I.~A.}\ \bibnamefont
  {Iakovlev}}, \bibinfo {author} {\bibfnamefont {Y.}~\bibnamefont {Li}},
  \bibinfo {author} {\bibfnamefont {A.}~\bibnamefont {W{\"o}rl}}, \bibinfo
  {author} {\bibfnamefont {Y.}~\bibnamefont {Tokiwa}}, \bibinfo {author}
  {\bibfnamefont {L.}~\bibnamefont {Ling}}, \bibinfo {author} {\bibfnamefont
  {Q.}~\bibnamefont {Zhang}}, \bibinfo {author} {\bibfnamefont {V.~V.}\
  \bibnamefont {Mazurenko}}, \bibinfo {author} {\bibfnamefont {P.}~\bibnamefont
  {Gegenwart}}, \ and\ \bibinfo {author} {\bibfnamefont {A.~A.}\ \bibnamefont
  {Tsirlin}},\ }\bibfield  {title} {\emph {\bibinfo {title} {Field evolution of
  the spin-liquid candidate {$YbMgGaO_4$}},\ }}\href@noop {} {\bibfield
  {journal} {\bibinfo  {journal} {arXiv preprint arXiv:2006.09775}\ } (\bibinfo
  {year} {2020})}\BibitemShut {NoStop}%
\bibitem [{\citenamefont {Ding}\ \emph {et~al.}(2020)\citenamefont {Ding},
  \citenamefont {Zhu}, \citenamefont {Zhang}, \citenamefont {Tan},
  \citenamefont {Yang}, \citenamefont {MacLaughlin},\ and\ \citenamefont
  {Shu}}]{ding2020persistent}%
  \BibitemOpen
  \bibfield  {author} {\bibinfo {author} {\bibfnamefont {Z.}~\bibnamefont
  {Ding}}, \bibinfo {author} {\bibfnamefont {Z.}~\bibnamefont {Zhu}}, \bibinfo
  {author} {\bibfnamefont {J.}~\bibnamefont {Zhang}}, \bibinfo {author}
  {\bibfnamefont {C.}~\bibnamefont {Tan}}, \bibinfo {author} {\bibfnamefont
  {Y.}~\bibnamefont {Yang}}, \bibinfo {author} {\bibfnamefont {D.~E.}\
  \bibnamefont {MacLaughlin}}, \ and\ \bibinfo {author} {\bibfnamefont
  {L.}~\bibnamefont {Shu}},\ }\bibfield  {title} {\emph {\bibinfo {title}
  {Persistent spin dynamics and absence of spin freezing in the $h-t$ phase
  diagram of the two-dimensional triangular antiferromagnet ${YbMgGaO}_{4}$},\
  }}\href {\doibase 10.1103/PhysRevB.102.014428} {\bibfield  {journal}
  {\bibinfo  {journal} {Phys. Rev. B}\ }\textbf {\bibinfo {volume} {102}},\
  \bibinfo {pages} {014428} (\bibinfo {year} {2020})}\BibitemShut {NoStop}%
\bibitem [{\citenamefont {Cevallos}\ \emph {et~al.}(2018)\citenamefont
  {Cevallos}, \citenamefont {Stolze},\ and\ \citenamefont
  {Cava}}]{cevallos2018structural}%
  \BibitemOpen
  \bibfield  {author} {\bibinfo {author} {\bibfnamefont {F.~A.}\ \bibnamefont
  {Cevallos}}, \bibinfo {author} {\bibfnamefont {K.}~\bibnamefont {Stolze}}, \
  and\ \bibinfo {author} {\bibfnamefont {R.~J.}\ \bibnamefont {Cava}},\
  }\bibfield  {title} {\emph {\bibinfo {title} {Structural disorder and
  elementary magnetic properties of triangular lattice {$ErMgGaO_4$} single
  crystals},\ }}\href@noop {} {\bibfield  {journal} {\bibinfo  {journal} {Solid
  State Commun.}\ }\textbf {\bibinfo {volume} {276}},\ \bibinfo {pages} {5}
  (\bibinfo {year} {2018})}\BibitemShut {NoStop}%
\bibitem [{\citenamefont {Xu}\ \emph {et~al.}(2019)\citenamefont {Xu},
  \citenamefont {Jiang}, \citenamefont {Cong},\ and\ \citenamefont
  {Yang}}]{xu2019structure}%
  \BibitemOpen
  \bibfield  {author} {\bibinfo {author} {\bibfnamefont {C.}~\bibnamefont
  {Xu}}, \bibinfo {author} {\bibfnamefont {P.}~\bibnamefont {Jiang}}, \bibinfo
  {author} {\bibfnamefont {R.}~\bibnamefont {Cong}}, \ and\ \bibinfo {author}
  {\bibfnamefont {T.}~\bibnamefont {Yang}},\ }\bibfield  {title} {\emph
  {\bibinfo {title} {Structure investigation of {$InGaZn_{1-x}Cu_{x}O_4$}
  (x=0-1) and magnetic property of {$InGaCuO_4$}},\ }}\href@noop {} {\bibfield
  {journal} {\bibinfo  {journal} {J. Solid State Chem.}\ }\textbf {\bibinfo
  {volume} {274}},\ \bibinfo {pages} {303} (\bibinfo {year}
  {2019})}\BibitemShut {NoStop}%
\bibitem [{\citenamefont {Shen}\ \emph {et~al.}(2019)\citenamefont {Shen},
  \citenamefont {Liu}, \citenamefont {Qin}, \citenamefont {Shen}, \citenamefont
  {Li}, \citenamefont {Bewley}, \citenamefont {Schneidewind}, \citenamefont
  {Chen},\ and\ \citenamefont {Zhao}}]{shen2019intertwined}%
  \BibitemOpen
  \bibfield  {author} {\bibinfo {author} {\bibfnamefont {Y.}~\bibnamefont
  {Shen}}, \bibinfo {author} {\bibfnamefont {C.}~\bibnamefont {Liu}}, \bibinfo
  {author} {\bibfnamefont {Y.}~\bibnamefont {Qin}}, \bibinfo {author}
  {\bibfnamefont {S.}~\bibnamefont {Shen}}, \bibinfo {author} {\bibfnamefont
  {Y.-D.}\ \bibnamefont {Li}}, \bibinfo {author} {\bibfnamefont
  {R.}~\bibnamefont {Bewley}}, \bibinfo {author} {\bibfnamefont
  {A.}~\bibnamefont {Schneidewind}}, \bibinfo {author} {\bibfnamefont
  {G.}~\bibnamefont {Chen}}, \ and\ \bibinfo {author} {\bibfnamefont
  {J.}~\bibnamefont {Zhao}},\ }\bibfield  {title} {\emph {\bibinfo {title}
  {Intertwined dipolar and multipolar order in the triangular-lattice magnet
  {$TmMgGaO_4$}},\ }}\href@noop {} {\bibfield  {journal} {\bibinfo  {journal}
  {Nat. Commun.}\ }\textbf {\bibinfo {volume} {10}},\ \bibinfo {pages} {1}
  (\bibinfo {year} {2019})}\BibitemShut {NoStop}%
\bibitem [{\citenamefont {Ma}\ \emph {et~al.}(2018{\natexlab{b}})\citenamefont
  {Ma}, \citenamefont {Ran}, \citenamefont {Wang}, \citenamefont {Bao},
  \citenamefont {Cai}, \citenamefont {Li},\ and\ \citenamefont
  {Wen}}]{ma2018recent}%
  \BibitemOpen
  \bibfield  {author} {\bibinfo {author} {\bibfnamefont {Z.}~\bibnamefont
  {Ma}}, \bibinfo {author} {\bibfnamefont {K.}~\bibnamefont {Ran}}, \bibinfo
  {author} {\bibfnamefont {J.}~\bibnamefont {Wang}}, \bibinfo {author}
  {\bibfnamefont {S.}~\bibnamefont {Bao}}, \bibinfo {author} {\bibfnamefont
  {Z.}~\bibnamefont {Cai}}, \bibinfo {author} {\bibfnamefont {S.}~\bibnamefont
  {Li}}, \ and\ \bibinfo {author} {\bibfnamefont {J.}~\bibnamefont {Wen}},\
  }\bibfield  {title} {\emph {\bibinfo {title} {Recent progress on
  magnetic-field studies on quantum-spin-liquid candidates},\ }}\href@noop {}
  {\bibfield  {journal} {\bibinfo  {journal} {Chin. Phys. B}\ }\textbf
  {\bibinfo {volume} {27}},\ \bibinfo {pages} {106101} (\bibinfo {year}
  {2018}{\natexlab{b}})}\BibitemShut {NoStop}%
\bibitem [{\citenamefont {Kasahara}\ \emph {et~al.}(2018)\citenamefont
  {Kasahara}, \citenamefont {Ohnishi}, \citenamefont {Mizukami}, \citenamefont
  {Tanaka}, \citenamefont {Ma}, \citenamefont {Sugii}, \citenamefont {Kurita},
  \citenamefont {Tanaka}, \citenamefont {Nasu}, \citenamefont {Motome} \emph
  {et~al.}}]{kasahara2018majorana}%
  \BibitemOpen
  \bibfield  {author} {\bibinfo {author} {\bibfnamefont {Y.}~\bibnamefont
  {Kasahara}}, \bibinfo {author} {\bibfnamefont {T.}~\bibnamefont {Ohnishi}},
  \bibinfo {author} {\bibfnamefont {Y.}~\bibnamefont {Mizukami}}, \bibinfo
  {author} {\bibfnamefont {O.}~\bibnamefont {Tanaka}}, \bibinfo {author}
  {\bibfnamefont {S.}~\bibnamefont {Ma}}, \bibinfo {author} {\bibfnamefont
  {K.}~\bibnamefont {Sugii}}, \bibinfo {author} {\bibfnamefont
  {N.}~\bibnamefont {Kurita}}, \bibinfo {author} {\bibfnamefont
  {H.}~\bibnamefont {Tanaka}}, \bibinfo {author} {\bibfnamefont
  {J.}~\bibnamefont {Nasu}}, \bibinfo {author} {\bibfnamefont {Y.}~\bibnamefont
  {Motome}},  \emph {et~al.},\ }\bibfield  {title} {\emph {\bibinfo {title}
  {Majorana quantization and half-integer thermal quantum {H}all effect in a
  {K}itaev spin liquid},\ }}\href@noop {} {\bibfield  {journal} {\bibinfo
  {journal} {Nature}\ }\textbf {\bibinfo {volume} {559}},\ \bibinfo {pages}
  {227} (\bibinfo {year} {2018})}\BibitemShut {NoStop}%
\bibitem [{\citenamefont {Xing}\ \emph
  {et~al.}(2019{\natexlab{b}})\citenamefont {Xing}, \citenamefont {Sanjeewa},
  \citenamefont {Kim}, \citenamefont {Stewart}, \citenamefont {Podlesnyak},\
  and\ \citenamefont {Sefat}}]{xing2019field}%
  \BibitemOpen
  \bibfield  {author} {\bibinfo {author} {\bibfnamefont {J.}~\bibnamefont
  {Xing}}, \bibinfo {author} {\bibfnamefont {L.~D.}\ \bibnamefont {Sanjeewa}},
  \bibinfo {author} {\bibfnamefont {J.}~\bibnamefont {Kim}}, \bibinfo {author}
  {\bibfnamefont {G.}~\bibnamefont {Stewart}}, \bibinfo {author} {\bibfnamefont
  {A.}~\bibnamefont {Podlesnyak}}, \ and\ \bibinfo {author} {\bibfnamefont
  {A.~S.}\ \bibnamefont {Sefat}},\ }\bibfield  {title} {\emph {\bibinfo {title}
  {Field-induced magnetic transition and spin fluctuations in the quantum
  spin-liquid candidate $csybse_2$},\ }}\href@noop {} {\bibfield  {journal}
  {\bibinfo  {journal} {Phys. Rev. B}\ }\textbf {\bibinfo {volume} {100}},\
  \bibinfo {pages} {220407} (\bibinfo {year} {2019}{\natexlab{b}})}\BibitemShut
  {NoStop}%
\bibitem [{\citenamefont {Shi}\ \emph {et~al.}(2019)\citenamefont {Shi},
  \citenamefont {Steinhardt}, \citenamefont {Graf}, \citenamefont {Corboz},
  \citenamefont {Weickert}, \citenamefont {Harrison}, \citenamefont {Jaime},
  \citenamefont {Marjerrison}, \citenamefont {Dabkowska}, \citenamefont {Mila}
  \emph {et~al.}}]{shi2019emergent}%
  \BibitemOpen
  \bibfield  {author} {\bibinfo {author} {\bibfnamefont {Z.}~\bibnamefont
  {Shi}}, \bibinfo {author} {\bibfnamefont {W.}~\bibnamefont {Steinhardt}},
  \bibinfo {author} {\bibfnamefont {D.}~\bibnamefont {Graf}}, \bibinfo {author}
  {\bibfnamefont {P.}~\bibnamefont {Corboz}}, \bibinfo {author} {\bibfnamefont
  {F.}~\bibnamefont {Weickert}}, \bibinfo {author} {\bibfnamefont
  {N.}~\bibnamefont {Harrison}}, \bibinfo {author} {\bibfnamefont
  {M.}~\bibnamefont {Jaime}}, \bibinfo {author} {\bibfnamefont
  {C.}~\bibnamefont {Marjerrison}}, \bibinfo {author} {\bibfnamefont {H.~A.}\
  \bibnamefont {Dabkowska}}, \bibinfo {author} {\bibfnamefont {F.}~\bibnamefont
  {Mila}},  \emph {et~al.},\ }\bibfield  {title} {\emph {\bibinfo {title}
  {Emergent bound states and impurity pairs in chemically doped
  {S}hastry-{S}utherland system},\ }}\href@noop {} {\bibfield  {journal}
  {\bibinfo  {journal} {Nat. Commun.}\ }\textbf {\bibinfo {volume} {10}},\
  \bibinfo {pages} {1} (\bibinfo {year} {2019})}\BibitemShut {NoStop}%
\bibitem [{\citenamefont
  {Rodr{\'\i}guez-Carvajal}(1993)}]{rodriguez1993recent}%
  \BibitemOpen
  \bibfield  {author} {\bibinfo {author} {\bibfnamefont {J.}~\bibnamefont
  {Rodr{\'\i}guez-Carvajal}},\ }\bibfield  {title} {\emph {\bibinfo {title}
  {Recent advances in magnetic structure determination by neutron powder
  diffraction},\ }}\href@noop {} {\bibfield  {journal} {\bibinfo  {journal}
  {Physica B Condens. Matter}\ }\textbf {\bibinfo {volume} {192}},\ \bibinfo
  {pages} {55} (\bibinfo {year} {1993})}\BibitemShut {NoStop}%
\bibitem [{\citenamefont {Van~Degrift}(1975)}]{van1975tunnel}%
  \BibitemOpen
  \bibfield  {author} {\bibinfo {author} {\bibfnamefont {C.~T.}\ \bibnamefont
  {Van~Degrift}},\ }\bibfield  {title} {\emph {\bibinfo {title} {Tunnel diode
  oscillator for 0.001 ppm measurements at low temperatures},\ }}\href@noop {}
  {\bibfield  {journal} {\bibinfo  {journal} {Rev. Sci. Instrum.}\ }\textbf
  {\bibinfo {volume} {46}},\ \bibinfo {pages} {599} (\bibinfo {year}
  {1975})}\BibitemShut {NoStop}%
\bibitem [{\citenamefont {Han}\ \emph {et~al.}(2012)\citenamefont {Han},
  \citenamefont {Helton}, \citenamefont {Chu}, \citenamefont {Nocera},
  \citenamefont {Rodriguez-Rivera}, \citenamefont {Broholm},\ and\
  \citenamefont {Lee}}]{han2012fractionalized}%
  \BibitemOpen
  \bibfield  {author} {\bibinfo {author} {\bibfnamefont {T.-H.}\ \bibnamefont
  {Han}}, \bibinfo {author} {\bibfnamefont {J.~S.}\ \bibnamefont {Helton}},
  \bibinfo {author} {\bibfnamefont {S.}~\bibnamefont {Chu}}, \bibinfo {author}
  {\bibfnamefont {D.~G.}\ \bibnamefont {Nocera}}, \bibinfo {author}
  {\bibfnamefont {J.~A.}\ \bibnamefont {Rodriguez-Rivera}}, \bibinfo {author}
  {\bibfnamefont {C.}~\bibnamefont {Broholm}}, \ and\ \bibinfo {author}
  {\bibfnamefont {Y.~S.}\ \bibnamefont {Lee}},\ }\bibfield  {title} {\emph
  {\bibinfo {title} {Fractionalized excitations in the spin-liquid state of a
  kagome-lattice antiferromagnet},\ }}\href@noop {} {\bibfield  {journal}
  {\bibinfo  {journal} {Nature}\ }\textbf {\bibinfo {volume} {492}},\ \bibinfo
  {pages} {406} (\bibinfo {year} {2012})}\BibitemShut {NoStop}%
\bibitem [{\citenamefont {Rosenkranz}\ and\ \citenamefont
  {Osborn}(2008)}]{rosenkranz2008corelli}%
  \BibitemOpen
  \bibfield  {author} {\bibinfo {author} {\bibfnamefont {S.}~\bibnamefont
  {Rosenkranz}}\ and\ \bibinfo {author} {\bibfnamefont {R.}~\bibnamefont
  {Osborn}},\ }\bibfield  {title} {\emph {\bibinfo {title} {Corelli: Efficient
  single crystal diffraction with elastic discrimination},\ }}\href@noop {}
  {\bibfield  {journal} {\bibinfo  {journal} {Pramana}\ }\textbf {\bibinfo
  {volume} {71}},\ \bibinfo {pages} {705} (\bibinfo {year} {2008})}\BibitemShut
  {NoStop}%
\bibitem [{\citenamefont {Michels-Clark}\ \emph {et~al.}(2016)\citenamefont
  {Michels-Clark}, \citenamefont {Savici}, \citenamefont {Lynch}, \citenamefont
  {Wang},\ and\ \citenamefont {Hoffmann}}]{michels2016expanding}%
  \BibitemOpen
  \bibfield  {author} {\bibinfo {author} {\bibfnamefont {T.~M.}\ \bibnamefont
  {Michels-Clark}}, \bibinfo {author} {\bibfnamefont {A.~T.}\ \bibnamefont
  {Savici}}, \bibinfo {author} {\bibfnamefont {V.~E.}\ \bibnamefont {Lynch}},
  \bibinfo {author} {\bibfnamefont {X.}~\bibnamefont {Wang}}, \ and\ \bibinfo
  {author} {\bibfnamefont {C.~M.}\ \bibnamefont {Hoffmann}},\ }\bibfield
  {title} {\emph {\bibinfo {title} {Expanding {L}orentz and spectrum
  corrections to large volumes of reciprocal space for single-crystal
  time-of-flight neutron diffraction},\ }}\href@noop {} {\bibfield  {journal}
  {\bibinfo  {journal} {J. Appl. Crystallogr.}\ }\textbf {\bibinfo {volume}
  {49}},\ \bibinfo {pages} {497} (\bibinfo {year} {2016})}\BibitemShut
  {NoStop}%
\bibitem [{\citenamefont {Chubukov}\ and\ \citenamefont
  {Golosov}(1991)}]{chubukov1991quantum}%
  \BibitemOpen
  \bibfield  {author} {\bibinfo {author} {\bibfnamefont {A.}~\bibnamefont
  {Chubukov}}\ and\ \bibinfo {author} {\bibfnamefont {D.}~\bibnamefont
  {Golosov}},\ }\bibfield  {title} {\emph {\bibinfo {title} {Quantum theory of
  an antiferromagnet on a triangular lattice in a magnetic field},\
  }}\href@noop {} {\bibfield  {journal} {\bibinfo  {journal} {J. Condens.
  Matter Phys.}\ }\textbf {\bibinfo {volume} {3}},\ \bibinfo {pages} {69}
  (\bibinfo {year} {1991})}\BibitemShut {NoStop}%
\bibitem [{\citenamefont {Li}\ \emph {et~al.}(2016{\natexlab{b}})\citenamefont
  {Li}, \citenamefont {Wang},\ and\ \citenamefont {Chen}}]{li2016anisotropic}%
  \BibitemOpen
  \bibfield  {author} {\bibinfo {author} {\bibfnamefont {Y.-D.}\ \bibnamefont
  {Li}}, \bibinfo {author} {\bibfnamefont {X.}~\bibnamefont {Wang}}, \ and\
  \bibinfo {author} {\bibfnamefont {G.}~\bibnamefont {Chen}},\ }\bibfield
  {title} {\emph {\bibinfo {title} {Anisotropic spin model of strong
  spin-orbit-coupled triangular antiferromagnets},\ }}\href@noop {} {\bibfield
  {journal} {\bibinfo  {journal} {Phys. Rev. B}\ }\textbf {\bibinfo {volume}
  {94}},\ \bibinfo {pages} {035107} (\bibinfo {year}
  {2016}{\natexlab{b}})}\BibitemShut {NoStop}%
\bibitem [{\citenamefont {Samarakoon}\ \emph {et~al.}(2019)\citenamefont
  {Samarakoon}, \citenamefont {Barros}, \citenamefont {Li}, \citenamefont
  {Eisenbach}, \citenamefont {Zhang}, \citenamefont {Ye}, \citenamefont {Dun},
  \citenamefont {Zhou}, \citenamefont {Grigera}, \citenamefont {Batista} \emph
  {et~al.}}]{samarakoon2019machine}%
  \BibitemOpen
  \bibfield  {author} {\bibinfo {author} {\bibfnamefont {A.~M.}\ \bibnamefont
  {Samarakoon}}, \bibinfo {author} {\bibfnamefont {K.}~\bibnamefont {Barros}},
  \bibinfo {author} {\bibfnamefont {Y.~W.}\ \bibnamefont {Li}}, \bibinfo
  {author} {\bibfnamefont {M.}~\bibnamefont {Eisenbach}}, \bibinfo {author}
  {\bibfnamefont {Q.}~\bibnamefont {Zhang}}, \bibinfo {author} {\bibfnamefont
  {F.}~\bibnamefont {Ye}}, \bibinfo {author} {\bibfnamefont {Z.}~\bibnamefont
  {Dun}}, \bibinfo {author} {\bibfnamefont {H.}~\bibnamefont {Zhou}}, \bibinfo
  {author} {\bibfnamefont {S.~A.}\ \bibnamefont {Grigera}}, \bibinfo {author}
  {\bibfnamefont {C.~D.}\ \bibnamefont {Batista}},  \emph {et~al.},\ }\bibfield
   {title} {\emph {\bibinfo {title} {Machine learning assisted insight to spin
  ice ${Dy_2Ti_2O_7}$},\ }}\href@noop {} {\bibfield  {journal} {\bibinfo
  {journal} {arXiv preprint arXiv:1906.11275}\ } (\bibinfo {year}
  {2019})}\BibitemShut {NoStop}%
\bibitem [{\citenamefont {Samarakoon}\ \emph {et~al.}(2017)\citenamefont
  {Samarakoon}, \citenamefont {Banerjee}, \citenamefont {Zhang}, \citenamefont
  {Kamiya}, \citenamefont {Nagler}, \citenamefont {Tennant}, \citenamefont
  {Lee},\ and\ \citenamefont {Batista}}]{samarakoon2017comprehensive}%
  \BibitemOpen
  \bibfield  {author} {\bibinfo {author} {\bibfnamefont {A.}~\bibnamefont
  {Samarakoon}}, \bibinfo {author} {\bibfnamefont {A.}~\bibnamefont
  {Banerjee}}, \bibinfo {author} {\bibfnamefont {S.-S.}\ \bibnamefont {Zhang}},
  \bibinfo {author} {\bibfnamefont {Y.}~\bibnamefont {Kamiya}}, \bibinfo
  {author} {\bibfnamefont {S.}~\bibnamefont {Nagler}}, \bibinfo {author}
  {\bibfnamefont {D.}~\bibnamefont {Tennant}}, \bibinfo {author} {\bibfnamefont
  {S.-H.}\ \bibnamefont {Lee}}, \ and\ \bibinfo {author} {\bibfnamefont
  {C.}~\bibnamefont {Batista}},\ }\bibfield  {title} {\emph {\bibinfo {title}
  {Comprehensive study of the dynamics of a classical {K}itaev spin liquid},\
  }}\href@noop {} {\bibfield  {journal} {\bibinfo  {journal} {Phys. Rev. B}\
  }\textbf {\bibinfo {volume} {96}},\ \bibinfo {pages} {134408} (\bibinfo
  {year} {2017})}\BibitemShut {NoStop}%
\bibitem [{\citenamefont {Kamiya}\ \emph {et~al.}(2018)\citenamefont {Kamiya},
  \citenamefont {Ge}, \citenamefont {Hong}, \citenamefont {Qiu}, \citenamefont
  {Quintero-Castro}, \citenamefont {Lu}, \citenamefont {Cao}, \citenamefont
  {Matsuda}, \citenamefont {Choi}, \citenamefont {Batista} \emph
  {et~al.}}]{kamiya2018nature}%
  \BibitemOpen
  \bibfield  {author} {\bibinfo {author} {\bibfnamefont {Y.}~\bibnamefont
  {Kamiya}}, \bibinfo {author} {\bibfnamefont {L.}~\bibnamefont {Ge}}, \bibinfo
  {author} {\bibfnamefont {T.}~\bibnamefont {Hong}}, \bibinfo {author}
  {\bibfnamefont {Y.}~\bibnamefont {Qiu}}, \bibinfo {author} {\bibfnamefont
  {D.}~\bibnamefont {Quintero-Castro}}, \bibinfo {author} {\bibfnamefont
  {Z.}~\bibnamefont {Lu}}, \bibinfo {author} {\bibfnamefont {H.}~\bibnamefont
  {Cao}}, \bibinfo {author} {\bibfnamefont {M.}~\bibnamefont {Matsuda}},
  \bibinfo {author} {\bibfnamefont {E.}~\bibnamefont {Choi}}, \bibinfo {author}
  {\bibfnamefont {C.}~\bibnamefont {Batista}},  \emph {et~al.},\ }\bibfield
  {title} {\emph {\bibinfo {title} {The nature of spin excitations in the
  one-third magnetization plateau phase of {Ba$_3$CoSb$_2$O$_9$}},\
  }}\href@noop {} {\bibfield  {journal} {\bibinfo  {journal} {Nat. Commun.}\
  }\textbf {\bibinfo {volume} {9}},\ \bibinfo {pages} {1} (\bibinfo {year}
  {2018})}\BibitemShut {NoStop}%
\bibitem [{\citenamefont {Nikuni}\ and\ \citenamefont
  {Jacobs}(1998)}]{nikuni1998quantum}%
  \BibitemOpen
  \bibfield  {author} {\bibinfo {author} {\bibfnamefont {T.}~\bibnamefont
  {Nikuni}}\ and\ \bibinfo {author} {\bibfnamefont {A.}~\bibnamefont
  {Jacobs}},\ }\bibfield  {title} {\emph {\bibinfo {title} {Quantum
  fluctuations in the incommensurate phase of {CsCuCl$_3$} in a transverse
  magnetic field},\ }}\href@noop {} {\bibfield  {journal} {\bibinfo  {journal}
  {Phys. Rev. B}\ }\textbf {\bibinfo {volume} {57}},\ \bibinfo {pages} {5205}
  (\bibinfo {year} {1998})}\BibitemShut {NoStop}%
\bibitem [{\citenamefont {Wen}\ \emph {et~al.}(2017)\citenamefont {Wen},
  \citenamefont {Koohpayeh}, \citenamefont {Ross}, \citenamefont {Trump},
  \citenamefont {McQueen}, \citenamefont {Kimura}, \citenamefont {Nakatsuji},
  \citenamefont {Qiu}, \citenamefont {Pajerowski}, \citenamefont {Copley} \emph
  {et~al.}}]{wen2017disordered}%
  \BibitemOpen
  \bibfield  {author} {\bibinfo {author} {\bibfnamefont {J.-J.}\ \bibnamefont
  {Wen}}, \bibinfo {author} {\bibfnamefont {S.}~\bibnamefont {Koohpayeh}},
  \bibinfo {author} {\bibfnamefont {K.}~\bibnamefont {Ross}}, \bibinfo {author}
  {\bibfnamefont {B.}~\bibnamefont {Trump}}, \bibinfo {author} {\bibfnamefont
  {T.}~\bibnamefont {McQueen}}, \bibinfo {author} {\bibfnamefont
  {K.}~\bibnamefont {Kimura}}, \bibinfo {author} {\bibfnamefont
  {S.}~\bibnamefont {Nakatsuji}}, \bibinfo {author} {\bibfnamefont
  {Y.}~\bibnamefont {Qiu}}, \bibinfo {author} {\bibfnamefont {D.}~\bibnamefont
  {Pajerowski}}, \bibinfo {author} {\bibfnamefont {J.}~\bibnamefont {Copley}},
  \emph {et~al.},\ }\bibfield  {title} {\emph {\bibinfo {title} {Disordered
  route to the {C}oulomb quantum spin liquid: random transverse fields on spin
  ice in {$Pr_2Zr_2O_7$}},\ }}\href@noop {} {\bibfield  {journal} {\bibinfo
  {journal} {Phys. Rev. Lett.}\ }\textbf {\bibinfo {volume} {118}},\ \bibinfo
  {pages} {107206} (\bibinfo {year} {2017})}\BibitemShut {NoStop}%
\bibitem [{\citenamefont {Furukawa}\ \emph {et~al.}(2015)\citenamefont
  {Furukawa}, \citenamefont {Miyagawa}, \citenamefont {Itou}, \citenamefont
  {Ito}, \citenamefont {Taniguchi}, \citenamefont {Saito}, \citenamefont
  {Iguchi}, \citenamefont {Sasaki},\ and\ \citenamefont
  {Kanoda}}]{furukawa2015quantum}%
  \BibitemOpen
  \bibfield  {author} {\bibinfo {author} {\bibfnamefont {T.}~\bibnamefont
  {Furukawa}}, \bibinfo {author} {\bibfnamefont {K.}~\bibnamefont {Miyagawa}},
  \bibinfo {author} {\bibfnamefont {T.}~\bibnamefont {Itou}}, \bibinfo {author}
  {\bibfnamefont {M.}~\bibnamefont {Ito}}, \bibinfo {author} {\bibfnamefont
  {H.}~\bibnamefont {Taniguchi}}, \bibinfo {author} {\bibfnamefont
  {M.}~\bibnamefont {Saito}}, \bibinfo {author} {\bibfnamefont
  {S.}~\bibnamefont {Iguchi}}, \bibinfo {author} {\bibfnamefont
  {T.}~\bibnamefont {Sasaki}}, \ and\ \bibinfo {author} {\bibfnamefont
  {K.}~\bibnamefont {Kanoda}},\ }\bibfield  {title} {\emph {\bibinfo {title}
  {Quantum spin liquid emerging from antiferromagnetic order by introducing
  disorder},\ }}\href@noop {} {\bibfield  {journal} {\bibinfo  {journal} {Phys.
  Rev. Lett.}\ }\textbf {\bibinfo {volume} {115}},\ \bibinfo {pages} {077001}
  (\bibinfo {year} {2015})}\BibitemShut {NoStop}%
\bibitem [{\citenamefont {Ranjith}\ \emph {et~al.}(2019)\citenamefont
  {Ranjith}, \citenamefont {Dmytriieva}, \citenamefont {Khim}, \citenamefont
  {Sichelschmidt}, \citenamefont {Luther}, \citenamefont {Ehlers},
  \citenamefont {Yasuoka}, \citenamefont {Wosnitza}, \citenamefont {Tsirlin},
  \citenamefont {K{\"u}hne} \emph {et~al.}}]{ranjith2019field}%
  \BibitemOpen
  \bibfield  {author} {\bibinfo {author} {\bibfnamefont {K.}~\bibnamefont
  {Ranjith}}, \bibinfo {author} {\bibfnamefont {D.}~\bibnamefont {Dmytriieva}},
  \bibinfo {author} {\bibfnamefont {S.}~\bibnamefont {Khim}}, \bibinfo {author}
  {\bibfnamefont {J.}~\bibnamefont {Sichelschmidt}}, \bibinfo {author}
  {\bibfnamefont {S.}~\bibnamefont {Luther}}, \bibinfo {author} {\bibfnamefont
  {D.}~\bibnamefont {Ehlers}}, \bibinfo {author} {\bibfnamefont
  {H.}~\bibnamefont {Yasuoka}}, \bibinfo {author} {\bibfnamefont
  {J.}~\bibnamefont {Wosnitza}}, \bibinfo {author} {\bibfnamefont {A.~A.}\
  \bibnamefont {Tsirlin}}, \bibinfo {author} {\bibfnamefont {H.}~\bibnamefont
  {K{\"u}hne}},  \emph {et~al.},\ }\bibfield  {title} {\emph {\bibinfo {title}
  {Field-induced instability of the quantum spin liquid ground state in the
  ${J}_{eff}= 1/2$ triangular-lattice compound {$NaYbO_2$}},\ }}\href@noop {}
  {\bibfield  {journal} {\bibinfo  {journal} {Phys. Rev. B}\ }\textbf {\bibinfo
  {volume} {99}},\ \bibinfo {pages} {180401} (\bibinfo {year}
  {2019})}\BibitemShut {NoStop}%
\bibitem [{\citenamefont {Mustonen}\ \emph {et~al.}(2018)\citenamefont
  {Mustonen}, \citenamefont {Vasala}, \citenamefont {Sadrollahi}, \citenamefont
  {Schmidt}, \citenamefont {Baines}, \citenamefont {Walker}, \citenamefont
  {Terasaki}, \citenamefont {Litterst}, \citenamefont {Baggio-Saitovitch},\
  and\ \citenamefont {Karppinen}}]{mustonen2018spin}%
  \BibitemOpen
  \bibfield  {author} {\bibinfo {author} {\bibfnamefont {O.}~\bibnamefont
  {Mustonen}}, \bibinfo {author} {\bibfnamefont {S.}~\bibnamefont {Vasala}},
  \bibinfo {author} {\bibfnamefont {E.}~\bibnamefont {Sadrollahi}}, \bibinfo
  {author} {\bibfnamefont {K.}~\bibnamefont {Schmidt}}, \bibinfo {author}
  {\bibfnamefont {C.}~\bibnamefont {Baines}}, \bibinfo {author} {\bibfnamefont
  {H.}~\bibnamefont {Walker}}, \bibinfo {author} {\bibfnamefont
  {I.}~\bibnamefont {Terasaki}}, \bibinfo {author} {\bibfnamefont
  {F.}~\bibnamefont {Litterst}}, \bibinfo {author} {\bibfnamefont
  {E.}~\bibnamefont {Baggio-Saitovitch}}, \ and\ \bibinfo {author}
  {\bibfnamefont {M.}~\bibnamefont {Karppinen}},\ }\bibfield  {title} {\emph
  {\bibinfo {title} {Spin-liquid-like state in a spin-1/2 square-lattice
  antiferromagnet perovskite induced by d10--d0 cation mixing},\ }}\href@noop
  {} {\bibfield  {journal} {\bibinfo  {journal} {Nat. Commun.}\ }\textbf
  {\bibinfo {volume} {9}},\ \bibinfo {pages} {1} (\bibinfo {year}
  {2018})}\BibitemShut {NoStop}%
\end{thebibliography}%
\end{document}